\documentclass[12pt]{article}
\usepackage{graphicx,psfrag,epsf,color}
\usepackage{hyperref}
\usepackage{multirow}
\usepackage{booktabs}
\usepackage{amsthm,amsmath,amssymb}
\usepackage{natbib}
\usepackage{mathtools}
\usepackage[linesnumbered, ruled, vlined]{algorithm2e}
\usepackage{centernot}

\newcommand{\toP}{\overset{P}{\longrightarrow}}
\newcommand{\toD}{\overset{\mathcal D}{\longrightarrow}}
\newcommand{\diag}{\operatorname{diag}}

\newtheorem{remark}{Remark}
\newtheorem{assumption}{Assumption}

\newtheorem{theorem}{Theorem}[section]
\newtheorem{lemma}{Lemma}[section] 
\newtheorem{proposition}{Proposition}[section]

\newcommand{\indep}{\rotatebox[origin=c]{90}{$\models$}}

\usepackage{tikz}
\tikzstyle{every node}=[font=\large,scale=1.2]
\hypersetup{
	colorlinks=true,
	linkcolor=red,
	urlcolor=blue,
	citecolor=blue
}

\title{A Generalized Fiducial Framework for Partially Identified Treatment Effects}
\author{
Yifan Cui\thanks{Center for Data Science, Zhejiang University, China. Correspondence to \href{mailto:cuiyf@zju.edu.cn}{cuiyf@zju.edu.cn}.}~
Jan Hannig\thanks{Department of Statistics and Operations Research, UNC-Chapel Hill, U.S.}~
}

\date{}

\begin{document}

\maketitle

\abstract{
Over the past two decades, there has been renewed interest in fiducial inference, a statistical framework originally proposed by R. A. Fisher in the 1930s. Existing contributions, however, have largely focused on point-identified models, where the parameter vector is uniquely determined by the joint distribution of the observables and the model assumptions. This paper develops a unified fiducial inference procedure for partially identified models, thereby extending the applicability of the framework to settings in which the identified set is non-degenerate. The acceptance rate of the proposed sampler naturally serves as a diagnostic: high values support the identifying assumptions, while near-zero values signal their potential violation. We establish a Bernstein-von Mises theorem for the fiducial distribution, which provides theoretical guarantees for the proposed sampler. As two leading examples, we provide uncertainty quantification for the instrumental variable model and the mediation model under a range of causal assumptions and target parameters. The proposed methodology is illustrated through extensive simulations and empirical applications.
}

\noindent {{\bf keywords:}
Acceptance Sampler, Fiducial Inference, Instrumental Variable, Mediation Analysis, Partial Identification, 
Uncertainty Quantification
}

\section{Introduction}

Fiducial inference, first introduced in a series of seminal papers by R. A. Fisher \citep{Fisher1922,Fisher1930,Fisher1933}, provides a framework for constructing probability distributions over parameters without imposing a prior. The central idea is to invert the data–parameter relationship, thereby inducing a distribution on the parameter space that reflects uncertainty about the unknown parameters based solely on the observed data.
Although fiducial inference itself has not achieved widespread popularity, its underlying ideas have been historically influential and continue to play an important role in contemporary statistical developments \citep{Efron1998}.
Recent developments, including generalized fiducial inference \citep{hannig2016generalized} and extended fiducial inference \citep{liang2025extended}, have broadened the applicability of this approach. These methods have been successfully applied in diverse domains such as psychology \citep{Liu2019,neupert2021bff}, forensic science \citep{hannig2019reported,hannig2022testing}, social network analysis \citep{su2022uncertainty}, and survival modeling \citep{cuihannig2019,cui2024unified,cui2026semiparametric}. See \cite{murph2023introduction,dawid2024fiducial,hannig2025fiducial} for recent reviews of fiducial inference, and \cite{XieSingh2013,martin2015inferential,schweder2016confidence,cui2025demystifying} for related frameworks on prior-free epistemic uncertainty quantification.

Despite recent advances in fiducial inference, its application to causal inference, particularly in settings with partially identified parameters, remains largely unexplored. In partially identified models, the observed data and structural assumptions do not uniquely determine the parameter of interest, but instead restrict it to a sharp identified set \citep{robins1989analysis,manski1990nonparametric,balke1994counterfactual,balke1997bounds,heckman1999local,imbens2004confidence,stoye2009more,chernozhukov2013intersection,molchanov2018random,cui2025policy}. 
To demonstrate the generality and practical utility of our proposed fiducial framework, we develop our methodology in a general formulation and then tailor it to two prototypical examples: instrumental variable (IV) models and causal mediation analysis. These settings represent two distinct yet complementary objectives, i.e., identifying overall causal effects under unmeasured confounding and dissecting underlying causal mechanisms along pathways. In both domains, point-identifying assumptions are frequently violated or untestable, rendering sharp partial identification and valid uncertainty quantification indispensable.

IV methods \citep{angrist1994,angrist1996identification} are a cornerstone for addressing unmeasured confounding in both observational studies and randomized trials with non-compliance. However, a valid instrument does not guarantee point identification of the average treatment effect without stringent auxiliary conditions, such as monotonicity or constant treatment effects. When these assumptions are relaxed or fail to hold, the causal effect is only partially identified \citep{robins1989analysis,balke1997bounds,imbens2004confidence,chernozhukov2007estimation,richardson2010analysis,chernozhukov2013intersection,richardson2015nonparametric,swanson2015bounding,swanson2018partial,kaido2019confidence,cui2021Individualized}. While characterizing the sharp bounds of partially identified IV models has received extensive attention, establishing tractable, prior-free inference and uncertainty quantification for these boundary estimands remains a critical challenge.

Causal mediation analysis \citep{robins1992identifiability,pearl2001direct,tchetgen2012semiparametric,vanderweele2015explanation} encounters analogous identification hurdles when decomposing total effects into natural direct and indirect mechanisms. Identifying these pathway-specific effects typically hinges on strong, untestable cross-world independence assumptions that are easily compromised by post-treatment confounding \citep{Avin2005identifiability}. To alleviate reliance on these restrictive conditions, an active literature has established non-parametric bounding strategies and partial identification results \citep{imai2010identification,tchetgen2014bounds,miles2017partial,steen2017medflex}. Nevertheless, existing methodologies primarily focus on deriving analytical bounds rather than providing a unified, distribution-free inferential framework. Developing exact, prior-free uncertainty quantification for these sharp mediation bounds remains an important open challenge.

In this paper, we develop a novel general fiducial framework for inference in partially identified statistical models, and apply it to the IV and mediation settings. We introduce an acceptance sampling algorithm to generate draws from the fiducial distribution of the identified set bounds. These fiducial samples can be used to construct point and interval estimates, e.g., the median fiducial draw as a point estimator, and empirical quantiles as confidence intervals for the lower and upper bounds, providing a unified approach to uncertainty quantification under partial identification. Moreover, the acceptance rate output by the acceptance sampler serves as a diagnostic for the compatibility of the data with the identifying assumptions.

This paper makes several contributions to the literature of causal inference and mathematical statistics. First, we develop a fiducial-based acceptance sampling algorithm for conducting inference on sharp bounds for a range of causal parameters under different identifying assumptions. Our approach not only preserves the sharp nonparametric nature of the frequentist bounds but also equips practitioners with an off-the-shelf, computationally coherent inferential machinery.
Second, we establish a Bernstein–von Mises theorem for the proposed procedure, thereby demonstrating that the resulting fiducial confidence intervals achieve asymptotically correct frequentist coverage for both the lower and upper bounds of the identified set. Third, the acceptance rate from our sampling algorithm has a natural interpretation as the fiducial probability that the observed data are consistent with the maintained identifying assumptions. This property yields an intuitive diagnostic: high acceptance rates lend credibility to the identifying assumptions, whereas rates approaching zero provide evidence against their validity.

The remainder of the article is organized as follows. In Section~\ref{sec:gfi}, we develop a unified fiducial inference framework for partially identified causal estimands, accompanied by an acceptance sampler and a diagnostic tool for model validation. In Section~\ref{sec:theory}, we establish a novel Bernstein-von Mises theorem for the proposed generalized fiducial distribution (GFD). 
In Sections~\ref{sec:pre} and \ref{sec:md}, we develop inferential tools for partially identified causal estimands in IV models and causal mediation analysis, respectively, which serve as two canonical special cases of our framework. Simulation studies are presented in Section~\ref{sec:simu}. Section~\ref{sec:real} describes two real data applications. 
The article concludes with a discussion in Section~\ref{sec:discuss}. Proofs and additional causal estimands and identifying assumptions under our framework are provided in the Supplementary Material.

\section{A unified fiducial framework}\label{sec:gfi}
\subsection{Preliminaries}
Across the various causal settings under consideration, including valid IV models, models relaxing instrument independence, causal mediation analysis, and population intervention effects, the computation of the bounds on the target estimand can be expressed within a unified fiducial framework. 

Let $O = (C, D, Y) \in \mathcal{C} \times \mathcal{D} \times \mathcal{Y}$ denote the vector of observed variables for an individual, where $C$ represents the randomized instrument (or exposure), $D$ denotes the treatment (or mediator), and $Y$ is the primary outcome. We keep the notation generic and set $C=Z, D=A$ in the IV models and $C=E, D=M$ in the mediation models. We consider the discrete setting with binary variables $\mathcal{C} = \mathcal{D} = \mathcal{Y} = \{0, 1\}$, and the empirical distribution of $O$ corresponds to cell counts or observed frequencies over the $|\mathcal{C} \times \mathcal{D} \times \mathcal{Y}| = 2^3$ observable configurations, denoted by $\mathbf{N}\in \mathbb{R}^8$. 

The observed data consist of $n$ realizations $\{O_i\}_{i=1}^n$. We denote by $n_{cdy}$ the number of times we have observed $C=c,D=d,Y=y$ and $q_{cdy}=P(D=d,Y=y|C=c)$  assuming that $0<\rho=P\left(  C=1\right)<1$.  Within each group $C=c$ let us assume that $\mathbf N_c=(n_{c00},n_{c01},n_{c10},n_{c11})$ follows Multinomial$(n_c,\mathbf q_c)$, with  $\mathbf q_c=(q_{c00},q_{c01},q_{c10}, q_{c11})^\top\in \Delta^3$, and $n_c=\sum_{dy} n_{cdy}$. This is weaker than assuming that the observed data $O_i$ are independent and identically distributed and includes cases where $C=1$ is assigned to a pre-selected proportion, e.g., 50\%, of observations.

Let $D_c$ be a person’s potential $D$ value under an intervention that sets $C$ to value $c$, and let $Y_{c,d}$ denote the potential outcome had, possibly contrary to fact, a person’s $C$ and $D$ been set to $c$ and $d$. We define the latent stratum probabilities
\begin{equation}\label{eq:gen_p_def}
p_{c,d_0 d_1,y_{00}y_{01}y_{10}y_{11}}= P(D_0=d_0,D_1=d_1,Y_{0,0}=y_{00},Y_{0,1}=y_{01},Y_{1,0}=y_{10},Y_{1,1}=y_{11}|C=c).
\end{equation}
Throughout, we assume the following consistency condition holds: $Y=Y_{C,D}$ and $D=D_C$. 
Consequently, the observed conditional probabilities are expressed in terms of these latent probabilities
\begin{equation}\label{eq:genQeq}
q_{cdy}=\sum_{d_c=d,y_{cd}=y} p_{c,d_0 d_1,y_{00}y_{01}y_{10}y_{11}}.
\end{equation}

Notice that for each $c=0,1$ there are potentially $K=64$ different latent conditional probabilities for a total of 128. However, model assumptions reduce the number of distinct non-zero probabilities by setting some to zero, constraining others to be equal, and may introduce additional constraints on the latent probabilities. Examples include IV independence, randomized treatment, exclusion restriction, monotonicity, and randomized mediators. As explained in Sections~\ref{sec:pre} and \ref{sec:md}, a valid binary IV model has only $K=16$ distinct probabilities independent of $c$, while causal mediation analysis needs $K = 64$ and introduces additional quadratic constraints.

To simplify the notation, we denote the vector of distinct conditional latent probabilities $\mathbf p=(\mathbf p_0,\mathbf p_1)\in\Delta^{K-1}\times\Delta^{K-1}$ and rewrite \eqref{eq:genQeq} together with additional model constraints as:
\begin{equation}
\label{eq:unified_const_q}
\begin{gathered}
 \mathbf{A}_{\text{obs}} \mathbf{p} = \mathbf{q}, \quad \mathbf{p} \ge \mathbf{0}, \quad  \mathbf{1}_K^\top \mathbf p_c = 1, \quad c \in \{0, 1\},\\
\mathbf{A}_{\text{eq}} \mathbf{p} = \mathbf{b}_{\text{eq}}, \quad
(\mathbf{W}_1 \mathbf{p}) \odot (\mathbf{W}_2 \mathbf{p}) - \mathbf{W}_3 \mathbf{p} = \mathbf{0}, 
\end{gathered}
\end{equation}
where $ \mathbf{A}_{\text{obs}}$ is determined by \eqref{eq:genQeq}. 
Next, define the following set of observable probabilities
\begin{equation}\label{eq:Fdef}
\mathcal F=\{\mathbf q\in\Delta^3\times\Delta^3;\mbox{ so that there exists $\mathbf p\in\Delta^{K-1}\times\Delta^{K-1}$ satisfying \eqref{eq:unified_const_q}}\}.
\end{equation}
Observe that only $\mathbf q\in\mathcal F$ are compatible with model assumptions.

Notice that the data can only provide us with information about $\mathbf q$. Therefore, for any true $\mathbf q\in\mathcal F$, we calculate the oracle lower and upper bounds by 
\begin{equation}
\label{eq:unified_opt_q}
\min_{\mathbf{p}} / \max_{\mathbf{p}} \quad  f(\mathbf p)=\frac{\boldsymbol{\alpha}^\top \mathbf{p} + \alpha_0}{\boldsymbol{\beta}^\top \mathbf{p} + \beta_0} \quad
\text{subject to constraints \eqref{eq:unified_const_q}}.
\end{equation}
Here
\begin{itemize}
    \item $\boldsymbol{\alpha}, \boldsymbol{\beta} \in \mathbb{R}^{2K}$ and $\alpha_0, \beta_0 \in \mathbb{R}$ define the objective function. For linear causal estimands, e.g., average treatment effect, natural direct effect, natural indirect effect, and population intervention indirect effect, $\boldsymbol{\beta} = \boldsymbol{0}$, $\beta_0 = 1$, and $\alpha_0 = 0$, simplifying the objective to $\boldsymbol{\alpha}^\top \mathbf{p}$. For fractional linear estimands, e.g., complier average treatment effect, average treatment effect on the treated, and nudge average treatment effect, $\boldsymbol{\alpha}^\top \mathbf{p} + \alpha_0$ and $\boldsymbol{\beta}^\top \mathbf{p} + \beta_0$ represent the numerator and denominator subpopulation terms, respectively.
    \item $\mathbf q=(\mathbf q_0, \mathbf q_1)\in\Delta^{3}\times\Delta^{3}$ is a vector of all observable conditional probabilities. 
    \item $\mathbf p=(\mathbf p_0, \mathbf p_1)\in\Delta^{K-1}\times\Delta^{K-1}$ is a vector of all latent conditional probabilities. 
    \item $\mathbf{A}_{\text{obs}} \in \{0, 1\}^{8 \times 2K}$ maps the latent probabilities to observable cell distributions.
    \item $\mathbf{A}_{\text{eq}} \in \{0, 1\}^{J \times 2K}$ and $\mathbf{b}_{\text{eq}} = \boldsymbol{0}$ encode linear structural restrictions, such as monotonicity or the new drug assumption, where $J$ is the number of constraints.
    \item $\mathbf{W}_1, \mathbf{W}_2, \mathbf{W}_3 \in \{0, 1\}^{R \times 2K}$ represent the $R$ nonlinear equality constraints arising from conditional or cross-world independence assumptions, with $\odot$ denoting the Hadamard product and $\mathbf{W}_3 = \mathbf{W}_1 \odot \mathbf{W}_2$. When no such constraints are imposed, $R = 0$.
\end{itemize}
When $R = 0$ and $\boldsymbol{\beta} = \boldsymbol{0}$, problem~\eqref{eq:unified_opt_q} reduces to a standard linear program; when $\boldsymbol{\beta} \neq \boldsymbol{0}$ with $R = 0$, it constitutes a linear-fractional program solvable via the Charnes-Cooper transformation; and when $R > 0$, it forms a quadratically constrained linear/fractional program. 

\begin{remark}
Many models, such as IV of Section~\ref{sec:pre} and mediation of Section~\ref{sec:md}, assume the independence of $C$ from $(D_c,Y_{cd})$, which leads to $\mathbf p_0 =\mathbf p_1$. While technically this can be encoded into the $\mathbf{A}_{\text{eq}} \mathbf{p} = \mathbf{b}_{\text{eq}}$ constraint, it would lead to an unnecessarily large optimization problem. Instead, without loss of generality, we commit slight abuse of notation, redefining $\mathbf p=\mathbf p_0$ and adjusting all the dimensions accordingly. 
\end{remark}

\subsection{GFDs for uncertainty quantification of multinomial distributions}
In order to assess uncertainty due to the randomness of the observed data, we derive a GFD for the probabilities $\mathbf p$ starting from GFDs for parameters of relevant multinomial distributions \citep{hannig2016generalized,lawrence2024new}. 

Fix $c\in\{0,1\}$, the GFD for $q_{cdy}$ is obtained by considering the random polygon determined by inequalities:
\begin{equation}\label{eq:MultinomialDSM}
V_{c00}^* \leq q_{c00},\quad
V_{c01}^* \leq q_{c01},\quad
V_{c10}^* \leq q_{c10},\quad
V_{c11}^* \leq q_{c11},
\end{equation}
where the random vector $(V_{c00}^*,V_{c01}^*,V_{c10}^*,V_{c11}^*)$ follows a distribution obtained by taking independent   Beta$(n_{cdy},1)$, $d,y\in\{0,1\}$ and truncating them to the event that a solution to \eqref{eq:MultinomialDSM} exists, i.e., $\mathcal B_c=\{V_{c00}^*+V_{c01}^*+V_{c10}^*+V_{c11}^*\leq 1\}$. Next, \cite{lawrence2024new} show that the distribution of 
$V_{c00}^*,\ldots,V_{c11}^*$ truncated to $\mathcal B_c$ is the same as the unconditional distribution of $V_{c00}^*,\ldots,V_{c11}^*$ where $(V_{c}^*,V_{c00}^*,\ldots,V_{c11}^*)$ follows Dirichlet$(1,n_{c00},\ldots,n_{c11})$. Furthermore, the random convex set determined by the inequalities in \eqref{eq:MultinomialDSM} is
\begin{multline}\label{eq:Sz}
  \mathcal S_c=\{\mathbf q_c\in\Delta^3;\mbox{ satisfying \eqref{eq:MultinomialDSM}}\}\\
=\operatorname{conv}\{(V_{c}^*+V_{c00}^*,V_{c01}^*,V_{c10}^*,V_{c11}^*),(V_{c00}^*,V_{c}^*+V_{c01}^*,V_{c10}^*,V_{c11}^*),\\(V_{c00}^*,V_{c01}^*,V_{c}^*+V_{c10}^*,V_{c11}^*),(V_{c00}^*,V_{c01}^*,V_{c10}^*,V_{c}^*+V_{c11}^*)\}.
\end{multline}
Any vector $q_{cdy}^*$ selected from the random polygon can be viewed as a sample from the GFD distribution for $q_{cdy}$. Moreover, for any set $\mathcal D$, $P^*(\mathcal S_c\cap \mathcal D\neq \emptyset)$ is called the plausibility of $\mathcal D$ and represents the fiducial belief that the set $\mathcal D$ is feasible given the observed data.

\subsection{GFD for partially identified parameters}
Let us now derive the GFD for our problem. We have observed two independent multinomial vectors, with parameters that are related to each other by \eqref{eq:genQeq}. Using the same arguments as in the development of the GFD for the multinomial distribution, we get $\mathbf V^*\leq \mathbf q$, where
$\mathbf V^*=(V^*_{000},\ldots,V^*_{111})$, with $V_{cdy}^*$ independent Beta$(n_{cdy},1)$ truncated to the set of 
\begin{equation}\label{eq:Vdef}
 \mathcal V=\{\mathbf v^*: \text{so that $\mathbf v^*\leq \mathbf q$ for some $\mathbf q$ satisfying \eqref{eq:unified_const_q}}\}.
\end{equation}

Thus, for a given fiducial draw $\mathbf {V}^* \in \mathcal V$, the lower and upper bounds, $l^*$ and $u^*$, are computed by solving:
\begin{equation}
\label{eq:unified_opt}
\begin{gathered}
\min_{\mathbf{p}} / \max_{\mathbf{p}}\  f(\mathbf p)=\frac{\boldsymbol{\alpha}^\top \mathbf{p} + \alpha_0}{\boldsymbol{\beta}^\top \mathbf{p} + \beta_0} \\
\text{s.t. } \mathbf{A}_{\text{obs}} \mathbf{p} \geq \mathbf V^*, \quad \mathbf{p} \ge \mathbf{0}, \quad  \mathbf{1}_K^\top \mathbf p_c = 1, \quad c \in \{0, 1\},\\
\mathbf{A}_{\text{eq}} \mathbf{p} = \mathbf{b}_{\text{eq}}, \quad
(\mathbf{W}_1 \mathbf{p}) \odot (\mathbf{W}_2 \mathbf{p}) - \mathbf{W}_3 \mathbf{p} = \mathbf{0},
\end{gathered}
\end{equation}
respectively. Sampling $\mathbf V^*\in\mathcal V$ introduces joint constraints on the parameters of the two multinomial distributions, and therefore there is no longer a closed form expression for the distribution of the random polygon. 
However, we can sample from it using a simple acceptance sampling algorithm.
To this end, we propose to generate  $(V_c^*,V_{c00}^*,V_{c01}^*,V_{c10}^*,V_{c11}^*)$ from Dirichlet$(1,n_{c00},n_{c01},n_{c10},n_{c11})$ independently for both $c=0,1$, and then accept the proposed combined $(V_{000}^*,\ldots,V_{111}^*)$ if there exist $\mathbf p$ that satisfy  \eqref{eq:unified_opt}. The details of the proposed acceptance sampler are described in Algorithm~\ref{alg:acceptanceate}.

\begin{algorithm}[h]
\SetAlgoLined
\caption{A causal acceptance sampler \label{alg:acceptanceate}}
\KwIn{Dataset $(C_i, D_i, Y_i)$, $n_{\text{mc}}$}
Attempts $\leftarrow$ 0\;
\For{$j \leftarrow 1$ \KwTo $n_{\text{mc}}$}{
    \Repeat{Feasible $(V_{000}^*,\ldots, V_{111}^*)$ found}{
        Attempts $\leftarrow$ Attempts $+\,1$\;
        Generate $(V_c^*,V_{c00}^*,V_{c01}^*,V_{c10}^*,V_{c11}^*)$ from Dirichlet$(1,n_{c00},n_{c01},n_{c10},n_{c11})$ independently for both $c=0,1$\;
        Check if  $\mathbf V^*\in\mathcal V$, where $\mathcal V$ is defined in \eqref{eq:Vdef}\;
    }
    $\mathbf V_j^* \leftarrow (V_{000}^*,\ldots, V_{111}^*)$\;
}
\Return Accepted fiducial samples $\mathbf V_j^*,\, j=1,\ldots,n_{\text{mc}}$ and acceptance rate $\frac{n_{\text{mc}}}{\text{Attempts}}$\;
\end{algorithm}

Surprisingly, the acceptance probability of our algorithm has a fiducial interpretation.
Note that Algorithm~\ref{alg:acceptanceate} accepts a proposed sample 
if and only if $(\mathcal S_0\times \mathcal S_1)\cap \mathcal F\neq\emptyset$, where $\mathcal F$ and $\mathcal S_c$ are defined in  \eqref{eq:Fdef} and \eqref{eq:Sz} respectively. 
Thus, the acceptance rate is an estimator of the fiducial probability $P^*((\mathcal S_0\times \mathcal S_1)\cap \mathcal F\neq\emptyset)$, which can be used to validate the feasibility of the model assumptions.
Consequently, the acceptance rate of the algorithm is an estimator of the fiducial probability of the observed data agreeing with the model assumptions.
 In other words, a high acceptance rate points toward high trust in the feasibility of the model assumptions, while an acceptance rate near zero suggests that the model assumptions are likely violated.

After generating fiducial samples $\mathbf V_j^*$, we propose Algorithm~\ref{alg:upperlowerate} to obtain upper and lower fiducial samples for a targeted causal parameter, e.g., the average treatment effect. Based on the fiducial samples, we construct pointwise confidence intervals by finding intervals of a given fiducial probability. In particular, a 95\% confidence interval is formed by taking the empirical 0.025 quantile of $l^*_j$ as a lower limit and the empirical 0.975 quantile of $u^*_j$ as an upper limit. We also propose using the pointwise median of $l^*_j$ and $u^*_j$ as point estimators for $l^*_j$ and $u^*_j$, respectively.

\begin{algorithm}[h] 
\SetAlgoLined
\caption{Generating upper and lower fiducial samples for a targeted estimand\label{alg:upperlowerate}}
\KwIn{The fiducial samples $\mathbf V_j^*$,    $j=1,\ldots,n_{\text{mc}}$.}
\For{$j = 1$ \textbf{to} $n_{\text{mc}}$}
{ 
\ShowLn Solve
$u^*_j = \max_{\Delta^{K-1}} f(\mathbf p)$ 
subject to constraints  \eqref{eq:unified_opt}\;
\ShowLn Solve
$l^*_j = \min_{\Delta^{K-1}} f(\mathbf p)$ 
subject to constraints \eqref{eq:unified_opt}\;
}
\Return  The upper and lower fiducial samples $u_j^*$ and $l_j^*$,    $j=1,\ldots,n_{\text{mc}}$.
\end{algorithm}

\subsection{Estimating the distribution of $C$}\label{sec:probability}
Some parameters of interest $f(\rho,\mathbf p)$, such as the average treatment effect on the treated, depend on $\rho=P(C=1)$ in addition to $\mathbf p$.  When $\rho$ is unknown, it will need to be estimated from the data. To be able to achieve this, we assume that the observed data $O_i$ are independent and identically distributed, which allows us to model the observed data as a single multinomial with $\bar q_{cdy}=P (D = d, Y =
y, C = c)$. To accommodate this, we need to update \eqref{eq:unified_const_q} by using 
\begin{equation}
\label{eq:updated_const_q}
\begin{gathered}
 \mathbf{\bar A}_{\text{obs}} \mathbf{p}=\mathbf{\bar q}, \quad \mathbf{p} \ge \mathbf{0}, \quad  \mathbf{1}_K^\top \mathbf p_c = 1, \quad c \in \{0, 1\},\\
\mathbf{A}_{\text{eq}} \mathbf{p} = \mathbf{b}_{\text{eq}}, \quad
(\mathbf{W}_1 \mathbf{p}) \odot (\mathbf{W}_2 \mathbf{p}) - \mathbf{W}_3 \mathbf{p} = \mathbf{0}, 
\end{gathered}
\end{equation}
where $\mathbf{\bar A}_{\text{obs}} = \mathbf W( \rho ) \mathbf{A}_{\text{obs}}$, $\mathbf W(\rho) = \begin{bmatrix} (1-\rho)\mathbf{I}_{4} & \mathbf{0} \\ \mathbf{0} & \rho\mathbf{I}_{4} \end{bmatrix}$, and $\rho\in(0,1)$. The feasible set $\bar{\mathcal V}$ is defined using the updated \eqref{eq:updated_const_q} instead of \eqref{eq:unified_const_q} in \eqref{eq:Vdef}.

Because the GFD is based on a single multinomial, sampling $\bar {\mathbf V}^* \in\bar{\mathcal V}$ only requires a single $\bar{\mathbf V}^*\sim $ Dirichlet$(1, \mathbf N)$ instead of two $\mathbf V^*_0,\mathbf V^*_1$. We then solve the updated optimization problem
\begin{equation}
\label{eq:updated_opt}
\begin{gathered}
\min_{\rho,\mathbf{p}} / \max_{\rho,\mathbf{p}}\  f(\rho, \mathbf p)\\
\text{s.t. } \mathbf{\bar A}_{\text{obs}} \mathbf{p} \geq \bar{\mathbf V}^*, \quad \mathbf{p} \ge \mathbf{0}, \quad  \mathbf{1}_K^\top \mathbf p_c = 1, \quad c \in \{0, 1\},\\
\mathbf{A}_{\text{eq}} \mathbf{p} = \mathbf{b}_{\text{eq}}, \quad
(\mathbf{W}_1 \mathbf{p}) \odot (\mathbf{W}_2 \mathbf{p}) - \mathbf{W}_3 \mathbf{p} = \mathbf{0}, \quad \rho\in(0,1).
\end{gathered}
\end{equation}

Note that even with $R=0$, one needs to solve a bilinear optimization problem. The bilinear coupling is driven solely by the scalar parameter $\rho \in (0, 1)$.
Moreover, adding the first and last four equations of $ \mathbf{\bar A}_{\text{obs}} \mathbf{p} \geq \bar{\mathbf V}^*$ leads to
\[
\bar V_{100}^* + \bar V_{101}^* + \bar V_{110}^* + \bar V_{111}^* 
\leq \rho \leq 1- (\bar V_{000}^* +\bar V_{001}^* +\bar V_{010}^* +\bar V_{011}^*)
=\bar V_{100}^* +\bar V_{101}^* + \bar V_{110}^* + \bar V_{111}^* + \bar V^*.
\]
Thus, the problem remains globally tractable via a line search over $\rho$.
Additional details are in Algorithm~\ref{alg:acceptancemd2}. 

\begin{algorithm}[h]
\SetAlgoLined
\caption{An acceptance sampler for the estimands involving $\rho$ \label{alg:acceptancemd2}}
\KwIn{Dataset $(C_i, D_i, Y_i)$, $n_{\text{mc}}$}
Attempts $\leftarrow$ 0\;
\For{$j \leftarrow 1$ \KwTo $n_{\text{mc}}$}{
    \Repeat{Feasible $(\bar{V}_{000}^*,\ldots, \bar{V}_{111}^*)\in\bar{\mathcal V}$ found}{
        Attempts $\leftarrow$ Attempts $+\,1$\;
        Generate $(\bar{V}^*,\bar{V}_{000}^*,\bar{V}_{001}^*,\bar{V}_{010}^*,\bar{V}_{011}^*,\bar{V}_{100}^*,\bar{V}_{101}^*,\bar{V}_{110}^*,\bar{V}_{111}^*)$ from Dirichlet$(1,n_{000},n_{001},n_{010},n_{011},n_{100},n_{101},n_{110},n_{111})$\;
    }
    $\bar{\mathbf V}_j^* \leftarrow (\bar{V}_{000}^*,\ldots, \bar{V}_{111}^*)$\;
}
\Return Accepted fiducial samples $\bar{\mathbf V}_j^*,\, j=1,\ldots,n_{\text{mc}}$ and acceptance rate $\frac{n_{\text{mc}}}{\text{Attempts}}$.
\end{algorithm}

\section{Bernstein-von Mises theorems}\label{sec:theory}

In this section, we present novel Bernstein-von Mises theorems for the GFDs. First, we provide the following lemma.

To fully appreciate the mode of convergence discussed below, notice that there are two sources of randomness present. First is the usual randomness of the
observed data, e.g., $\mathbf N$. The second source of randomness is the
fiducial distribution defined for each fixed data set. We indicate this randomness by using stars, e.g., $P^*, \mathbf V^*$. We prefer this notation to using a conditioning symbol, as the fiducial distribution was not obtained by Bayes' theorem. We denote convergence in distribution by $\toD$. The main results use convergence in distribution in probability. In particular, 
$X^*\toD X$ in probability if the L\'evy-Prokhorov distance $\nu(X^*,X)\toP 0$.

\begin{lemma}\label{lm:Dirichlet}
Let $\mathbf N\in \mathbb R^l$ be Multinomial$(n,\mathbf q)$, $\mathbf q\in\Delta^{l-1}$. Let $\mathbf V^*|\mathbf N=\mathbf n$ be Dirichlet$(1,\mathbf n)$. Then, as $n\to\infty$ 
  \begin{enumerate}
      \item $\sqrt{n}(\mathbf N/n-\mathbf q)\toD N(\mathbf 0,\Sigma)$, where $\Sigma=\diag(\mathbf q)-\mathbf{qq}^\top$.
      
      \item $\sqrt{n}\left(\mathbf V^*-\mathbf N^0/n\right)\toD N(\mathbf 0,\Sigma^0)$ in probability where 
      \[
       \mathbf N^0=\begin{pmatrix}
          0\\ \mathbf  N
      \end{pmatrix}
      \mbox{ and~ }
      \Sigma^0
      =\begin{pmatrix}
        0 & \mathbf 0^\top\\
        \mathbf 0 & \Sigma
        \end{pmatrix}.
    \]
  \end{enumerate}
\end{lemma}

The following theorem validates our assertion that the acceptance rate can be used to assert whether the model assumptions are satisfied. In what follows, $\mathcal F^\circ$ is the interior of $\mathcal F$ given by \eqref{eq:Fdef}. 
\begin{theorem}\label{thm:acceptance}
 Let $\mathbf N_c=(n_{c00},\ldots, n_{c11})$ be independent Multinomial$(n_c,\mathbf q_c)$, with  $\mathbf q_c=(q_{c00},q_{c01},q_{c10}, q_{c11})^\top\in \Delta^3$ and $c\in\{0,1\}$. 
 Given $\mathbf N_c=\mathbf n_c$, $c\in\{0,1\}$, let $\mathbf V^*=(V_{c}^*,V_{c00}^*,\ldots,V_{c11}^*)$ be independent Dirichlet$(1,\mathbf n_c)$, $c\in\{0,1\}$.  Then for $
 \mathbf q=(q_{000},\ldots,q_{111})$, as $n_0,n_1\to\infty$:
\begin{enumerate}
\item if $\mathbf q\in\mathcal F^\circ$, then $P^*((\mathcal S_0\times \mathcal S_1)\cap \mathcal F\neq\emptyset)\toP 1$;
\item if $\mathbf q\in\mathcal (\mathcal F^\complement)^\circ$, then 
 $P^*((\mathcal S_0\times \mathcal S_1)\cap \mathcal F\neq\emptyset)\toP 0$;
 \item if $\mathbf q\in \partial \mathcal F$ and $n_c/(n_0+n_1)\to\lambda_c\in(0,1)$, then 
 \[\liminf P(P^*((\mathcal S_0\times \mathcal S_1)\cap \mathcal F\neq\emptyset)\leq u)\geq u \mbox{ for all } u\in(0,1).
 \]
 \end{enumerate}
\end{theorem}

Note that the feasibility set $\mathcal F$ is a convex polytope. If $\mathbf q\in\partial\mathcal F$ is additionally in the interior of a face, 
then the inequality in \eqref{eq:acceptaceComp} becomes equality in the limit and the acceptance rate itself actually converges to the uniform distribution. Thus, large values of the acceptance rate indicate that we are safely within the region where the model assumptions are satisfied; small values mean that the model assumptions are likely violated, while values in between indicate we may be near the boundary.

Next, we prove the Bernstein-von Mises theorem for solutions of an optimization problem in Algorithm~\ref{alg:upperlowerate} or one of its variants. We will use the notation introduced in the previous theorem.
\begin{theorem}\label{thm:BvM2}
Let $\mathbf N_c$ be independent Multinomial$(n_c,\mathbf q_c)$, $c\in\{0,1\}$. 
Let the conditional distribution $\mathbf V_c^* | \mathbf N_c=\mathbf n_c$  be independent Dirichlet$(1,\mathbf n_c)$, $c\in\{0,1\}$.

   For each $\tilde{\mathbf q}\in\mathcal F$ define $g(\tilde{\mathbf q}) = \min_{\mathbf p} f(\mathbf p)$ subject to \eqref{eq:unified_const_q}.
   Assume that there is an open set $\mathcal Q$ such that $\mathbf q\in\mathcal Q\subset\mathcal F$, and the function $g(\tilde{\mathbf q})$ is continuously differentiable on $\mathcal Q$. Set  $n_c/(n_0+n_1)\to\lambda_c\in(0,1)$. Then as $n_0,n_1\to\infty$
  \begin{enumerate}
   \item $\sqrt{n_0+n_1}(g(\mathbf N_0/n_0,\mathbf N_1/n_1)-g(\mathbf q))\toD N(0, \sigma^2),$ where 
   \[
   \sigma^2=\sum_{c=0}^1 \lambda_c^{-1}\left(\sum_{d,y=0}^1  \left(\frac{\partial  g(\mathbf q)}{\partial q_{cdy}}\right)^2 q_{cdy} -  \left(\sum_{d,y=0}^1\frac{\partial g(\mathbf q)}{\partial q_{cdy}}  q_{cdy}\right)^2\right).\]

   \item Denote the random variable $G^*=\min_{\mathbf p} f(\mathbf p)$ subject to the constraints in \eqref{eq:unified_opt}, we have
   \[
     \sqrt{n_0+n_1}(G^*-g(\mathbf N_0/n_0,\mathbf N_1/n_1))\toD N(0, \sigma^2) \mbox{ in probability}.
   \]
   \end{enumerate}
\end{theorem}

Note that Theorem~\ref{thm:BvM2} also holds for $\min$ replaced with $\max$. The consequence of this result is that we have a correct coverage for confidence intervals for the theoretical lower and upper bounds on the quantity of interest. 

The assumption on differentiability of $g(\mathbf q)$ is satisfied if the objective function $f$ is continuously differentiable and the locations of non-zero entries in the minimizer $\mathbf p=(p_{0},\ldots,p_{K})$  are the same for all $\mathbf q\in\mathcal Q$.

Next we present a Bernstein-von Mises theorem for the fiducial distribution sampled from Algorithm~\ref{alg:acceptancemd2} when the estimand of interest involves $\rho=P(C=1)$. 

\begin{theorem}\label{thm:BvM2A}
Let $\mathbf N$ be independent Multinomial$(n,(1-\rho)\mathbf q_0,\rho\mathbf q_1)$. 
Let the conditional distribution $\mathbf V^* | \mathbf N=\mathbf n$  be independent Dirichlet$(1,\mathbf n_0,\mathbf n_1)$.

   For each $\tilde{\mathbf q}\in\mathcal F$ and $\tilde\rho\in(0,1)$, define $g(\tilde \rho,\tilde{\mathbf q}) = \min_{\mathbf p} f(\tilde\rho,\mathbf p)$ subject to \eqref{eq:updated_const_q}.
   Assume that there is an open set $\mathcal Q$ so that $(\rho,\mathbf q)\in\mathcal Q\subset [0,1]\times\mathcal F$, and the function $g(\tilde\rho,\tilde{\mathbf q})$ is continuously differentiable on $\mathcal Q$. Then as $n\to\infty$,
  \begin{enumerate}
   \item $\sqrt{n}(g(n_1/n,\mathbf N_0/n_0,\mathbf N_1/n_1)-g(\rho,\mathbf q))\toD N(0, \sigma^2),$ where 
   \begin{align*}
   \sigma^2=&\rho(1-\rho)\left(\frac{\partial g(\rho,\mathbf q)}{\partial \rho}\right)^2+ 
   \rho^{-1}\left(\sum_{d,y=0}^1  \left(\frac{\partial  g(\rho,\mathbf q)}{\partial q_{1dy}}\right)^2 q_{1dy} -  \left(\sum_{d,y=0}^1\frac{\partial g(\rho,\mathbf q)}{\partial q_{1dy}}  q_{1dy}\right)^2\right)\\
   &+(1-\rho)^{-1}\left(\sum_{d,y=0}^1  \left(\frac{\partial  g(\rho,\mathbf q)}{\partial q_{0dy}}\right)^2 q_{0dy} -  \left(\sum_{d,y=0}^1\frac{\partial g(\rho,\mathbf q)}{\partial q_{0dy}}  q_{0dy}\right)^2\right).
   \end{align*}

   \item Denote the random variable $G^*=\min_{\rho, \mathbf p} f(\rho,\mathbf p)$ subject to the constraints in \eqref{eq:updated_opt}, we have
   \[
     \sqrt{n}(G^*-g(n_1/n,\mathbf N_0/n_0,\mathbf N_1/n_1))\toD N(0, \sigma^2) \mbox{ in probability}.
   \]
   \end{enumerate}
\end{theorem}

By the above theorems, we have coverage guarantees for confidence intervals for the theoretical lower and upper bounds of the estimands of interest, respectively.

\section{Example 1: IV models}\label{sec:pre}
\subsection{IV model revisited}

In this section and the counterparts of the numerical sections, set $C=Z$ and $D=A$. Let $Y$ denote the outcome of interest and $A \in \{0,1\}$ be a binary treatment indicator. There might be unmeasured confounders of the effect of $A$ on $Y$ which are not known a priori. Suppose also that one has observed a binary IV $Z \in \{0,1\}$. 
We assume the following set of IV assumptions \citep{angrist1994,balke1997bounds,swanson2018partial}:

\begin{assumption}{(IV independence)}\label{asm:ivindep} $Z \indep \{Y_{a=0}, Y_{a=1}, A_{z=0}, A_{z=1}\}$.
\end{assumption}

\begin{assumption}{(Exclusion restriction)} \label{asm:er} $Y_{z,a}=Y_a$.
\end{assumption}


Assumption~\ref{asm:ivindep} ensures that the causal effect of $Z$ on $A$ and $Y$ is unconfounded. 
Assumption~\ref{asm:er} assumes an individual-level exclusion restriction, i.e., there can be no individual direct causal effect of $Z$ on $Y$ not mediated by $A$.
Figure~\ref{fig1} gives an illustration of a valid IV.

\vspace{0.6cm}
\begin{figure}[h]
\centering
\begin{tikzpicture}[state/.style={circle, draw, minimum size=1cm}]
  \def\Ax{0}
  \def\Ay{0}
  \def\offset{2.5}
  \def\Bx{\Ax+5}
  \def\By{\Ay}
  \node[state,shape=circle,draw=black] (Z) at (\Ax,\Ay) {$Z$};
  \node[state,shape=circle,draw=black] (Y) at (\Bx,\By) {$Y$};
  \node[state,shape=circle,draw=black] (A) at (\Bx-\offset,\By) {$A$};
  \node[state,shape=circle,draw=black,fill=black!20] (U) at (\Bx-1.25,\By-1.5) {$U$};

  \draw [-latex] (Z) to [bend left=0] (A);
  \draw [-latex] (A) to [bend left=0] (Y);
  \draw [-latex] (U) to [bend left=0] (A);
  \draw [-latex] (U) to [bend left=0] (Y);

\end{tikzpicture}
\caption{A directed acyclic graph (DAG) for an IV, where $U$ refers to unmeasured confounders.\label{fig1}}
\end{figure}

The IV assumptions impose constraints on the latent probabilities in \eqref{eq:gen_p_def}. First, Assumption~\ref{asm:ivindep} of IV independence gives $\mathbf p_0=\mathbf p_1$. Second, Assumption~\ref{asm:er} on the exclusion restriction implies that
$p_{z,d_0 d_1,y_{00}y_{01}y_{10}y_{11}}=0$ if either $y_{00}\neq y_{10}$ or $y_{01}\neq y_{11}$. Consequently, the vector $\mathbf p=\mathbf p_0$ only contains $K=16$ distinct latent strata probabilities $p_{a_0a_1,y_0y_1}$ that are needed to describe causal effects: 
\[p_{a_0a_1,y_0y_1}=P(A_0=a_0,A_1=a_1,Y_0=y_0,Y_1=y_1),\quad a_0,a_1,y_0,y_1\in\{0,1\}.\]
Naturally, the vector $\mathbf p=(p_{00,00},\ldots,p_{11,11})^\top\in \Delta^{15}$.
Table~\ref{table:main1} shows the partition of the population into 16 disjoint sets. In the Supplementary Material~\ref{sec:32numbers}, we partition the population into 32 disjoint sets to further relax Assumption \ref{asm:ivindep}.

\begin{table}[h]
\centering
\begin{tabular}{ccccccccc}
& Never-taker & Prob  & Complier  & Prob  & Defier & Prob & Always-taker & Prob\\
\hline
Doomed & 00,00 & $p_{00,00}$  & 01,00 & $p_{01,00}$ & 10,00 & $p_{10,00}$ & 11,00 & $p_{11,00}$ \\
Beneficial & 00,01 & $p_{00,01}$ & 01,01 & $p_{01,01}$ & 10,01 & $p_{10,01}$ & 11,01 & $p_{11,01}$ \\
Harmed & 00,10 & $p_{00,10}$  & 01,10 & $p_{01,10}$ & 10,10 & $p_{10,10}$ & 11,10 & $p_{11,10}$ \\
Immune & 00,11 & $p_{00,11}$  & 01,11 & $p_{01,11}$ & 10,11 & $p_{10,11}$ & 11,11 & $p_{11,11}$\\
\hline
\end{tabular}
\caption{A partition of the population into 16 disjoint sets.\label{table:main1}}
\end{table}

These probabilities are useful in describing many causal quantities of interest. For example, the celebrated average treatment effect (ATE) 
\begin{align*}
     E(Y_1-Y_0)=& P(Y_1=1)-P(Y_0=1)\\
     =& p_{00,01} + p_{01,01} + p_{10,01} + p_{11,01} - p_{00,10} - p_{01,10}- p_{10,10}- p_{11,10}.
\end{align*} 

\subsection{Fiducial bounds on average treatment effects}
We will derive fiducial-based stochastic bounds on the ATE in this section and consider other causal estimands under a variety of assumptions in the Supplementary Material~\ref{sec:estimand} and \ref{sec:assumption}. To this end, we first link these unobservable probabilities with probabilities of observable events in the following proposition.
\begin{proposition}\label{prop:qp}
Under Assumptions~\ref{asm:ivindep} and \ref{asm:er},
\begin{equation}\label{eq:q2p}
\begin{aligned}
    q_{000} &= p_{00,00} + p_{00,01} +  p_{01,00} + p_{01,01} \\
    q_{001} &= p_{00,10} + p_{00,11} + p_{01,10} + p_{01,11} \\
    q_{010} &= p_{10,00} + p_{10,10} + p_{11,00} + p_{11,10} \\
    q_{011} &= p_{10,01} + p_{10,11} + p_{11,01} + p_{11,11} \\
    q_{100} &= p_{00,00} + p_{00,01} + p_{10,00} + p_{10,01}\\
    q_{101} &= p_{00,10} + p_{00,11} + p_{10,10} + p_{10,11}\\
    q_{110} &= p_{01,00} + p_{01,10} + p_{11,00} + p_{11,10} \\
    q_{111} &= p_{01,01} + p_{01,11} + p_{11,01} + p_{11,11}.
\end{aligned}
\end{equation}
\end{proposition}

Note that in \eqref{eq:q2p}  the sum of the first four equations is one, as is the sum of the last four equations. The right-hand side of \eqref{eq:q2p} determines the matrix $\mathbf A_{\text{obs}}$ in \eqref{eq:unified_opt}, i.e., 
\begin{equation}
\mathbf{A}_{\text{obs}} = 
\begin{pmatrix}
\mathbf{B}_1   & \mathbf{B}_1 & \mathbf{0}  & \mathbf{0}   \\
\mathbf{B}_2    & \mathbf{B}_2 & \mathbf{0} & \mathbf{0}   \\
\mathbf{0}  & \mathbf{0}    & \mathbf{B}_3  & \mathbf{B}_3 \\
\mathbf{0}   & \mathbf{0}   & \mathbf{B}_4 & \mathbf{B}_4 \\
\mathbf{B}_1 & \mathbf{0}   & \mathbf{B}_1 & \mathbf{0}    \\
\mathbf{B}_2 & \mathbf{0}  & \mathbf{B}_2  & \mathbf{0}   \\
\mathbf{0}   & \mathbf{B}_3 & \mathbf{0} & \mathbf{B}_3    \\
\mathbf{0}   & \mathbf{B}_4 & \mathbf{0} & \mathbf{B}_4 
\end{pmatrix},
\end{equation}
where $\mathbf{B}_1 = (1, 1, 0, 0)$, $\mathbf{B}_2 = (0, 0, 1, 1)$, $\mathbf{B}_3 = (1, 0, 1, 0)$, $\mathbf{B}_4 = (0, 1, 0, 1)$, and $\mathbf{0} = (0, 0, 0, 0)$.

We observe two independent multinomial vectors, with parameters related by Proposition~\ref{prop:qp}. Thus, by combining \eqref{eq:q2p} and \eqref{eq:MultinomialDSM}, the joint GFD for this problem is sampled from the random polygon
\begin{equation}\label{eq:causalDSM}
    \begin{aligned}
V_{000}^* &\leq p_{00,00} + p_{00,01} + p_{01,00} + p_{01,01}\\
V_{001}^* &\leq p_{00,10} + p_{00,11} + p_{01,10} + p_{01,11} \\
V_{010}^* &\leq p_{10,00} + p_{10,10} + p_{11,00} + p_{11,10} \\
V_{011}^* &\leq p_{10,01} + p_{10,11} + p_{11,01} + p_{11,11} \\
V_{100}^* &\leq p_{00,00} + p_{00,01} + p_{10,00} + p_{10,01} \\
V_{101}^* &\leq p_{00,10} + p_{00,11} + p_{10,10} + p_{10,11} \\
V_{110}^* &\leq p_{01,00} + p_{01,10} + p_{11,00} + p_{11,10} \\
V_{111}^* &\leq p_{01,01} + p_{01,11} + p_{11,01} + p_{11,11},       
    \end{aligned}
\end{equation}
where again $V^*_{zay} \sim$ Beta$(n_{zay}, 1)$ are independent random variables truncated to the event that a solution $\mathbf p\in\Delta^{15}$ of  \eqref{eq:causalDSM} exists.

\begin{remark}
Importantly, the proposed fiducial bounds are asymptotically equivalent to the Balke-Pearl bounds \citep{balke1997bounds}, while providing valid uncertainty quantification in an intrinsic and automated fashion. While the classical Balke-Pearl framework characterizes the sharp identification region through linear programming over the latent response types, translating these bounds into practical statistical inference typically requires non-trivial secondary procedures, such as plug-in sample analogues coupled with ad-hoc boundary corrections or bootstrap calibrations to handle non-differentiabilities. In contrast, our fiducial formulation directly generates an entire epistemic distribution over the parameter space without relying on subjective priors. Consequently, the resulting fiducial credible intervals naturally reflect sampling variability, circumventing the need to derive complex nuisance parameters or asymptotic distribution approximations for piecewise-linear boundary functionals.
\end{remark}

\subsection{Extension to average treatment effect on the treated}
Estimands such as the average treatment effect on the treated $E[Y_1 - Y_0 \mid A=1]$ \citep{hernan2023causal} can be represented using $\mathbf p$ and $\rho=P(Z=1)$ as follows:
\begin{align*}
&[(p_{01,01}+p_{11,01}-p_{01,10}-p_{11,10})\rho+(p_{10,01}+p_{11,01}-p_{10,10}-p_{11,10})(1-\rho)]\\
&\times[(p_{01,00}+p_{01,01}+p_{01,10}+p_{01,11}+p_{11,00}+p_{11,01}+p_{11,10}+p_{11,11})\rho\\
&\ +(p_{10,00}+p_{10,01}+p_{10,10}+p_{10,11}+p_{11,00}+p_{11,01}+p_{11,10}+p_{11,11})(1-\rho)]^{-1},
\end{align*}
 where the derivation is in the Supplementary Material~\ref{sec:G2}.
 
 We concretize Section~\ref{sec:probability} as follows,
\begin{equation}\label{eq:causalDSMjoint}
    \begin{aligned}
\bar V_{000}^* &\leq \bar q_{000}=(p_{00,00} + p_{00,01} +  p_{01,00} + p_{01,01})(1-\rho) \\
\bar V_{001}^* &  \leq \bar q_{001} = (p_{00,10} + p_{00,11} + p_{01,10} + p_{01,11})(1-\rho) \\
\bar V_{010}^* &\leq \bar q_{010} = (p_{10,00} + p_{10,10} + p_{11,00} + p_{11,10})(1-\rho) \\
\bar V_{011}^* &\leq \bar q_{011} = (p_{10,01} + p_{10,11} + p_{11,01} + p_{11,11})(1-\rho) \\
\bar V_{100}^* &\leq \bar q_{100} = (p_{00,00} + p_{00,01} + p_{10,00} + p_{10,01})\rho \\
\bar V_{101}^* &\leq \bar q_{101} = (p_{00,10} + p_{00,11} + p_{10,10} + p_{10,11})\rho \\
\bar V_{110}^* &\leq \bar q_{110} = (p_{01,00} + p_{01,10} + p_{11,00} + p_{11,10})\rho \\
\bar V_{111}^* &\leq \bar q_{111} = (p_{01,01} + p_{01,11} + p_{11,01} + p_{11,11})\rho,       
    \end{aligned}
\end{equation}
where $\bar V^*_{zay} \sim$ Beta$(n_{zay}, 1)$ are independent random variables truncated to the event that a solution $\rho\in(0,1)$, $\mathbf p\in\Delta^{15}$ of  \eqref{eq:causalDSMjoint} exists. In particular, we generate  $(\bar V^*,\bar V_{000}^*,\bar V_{001}^*,\ldots \bar V_{110}^*,\bar V_{111}^*)$ from Dirichlet$(1,n_{000},n_{001},\ldots,n_{110},n_{111})$ and accept the proposed combined $(\bar V_{000}^*,\ldots,\bar V_{111}^*)$ if there exist $\rho, \mathbf p$ that satisfy  \eqref{eq:causalDSMjoint}.

\section{Example 2: mediation analysis}\label{sec:md}

\subsection{Mediation model revisited}
In this section, we extend the aforementioned fiducial idea to causal mediation analysis. We set $C=E$ and $D=M$ in the mediation analysis.
Our goal is to decompose the total effect of an exposure $E$ on an outcome $Y$ into two distinct pathways: the natural direct effect (NDE) and the natural indirect effect (NIE). See Figure~\ref{fig2} for a causal directed acyclic graph. 

\vspace{0.2cm}
\begin{figure}[h]
\centering
\begin{tikzpicture}[state/.style={circle, draw, minimum size=1cm}]
  \def\Ax{0}
  \def\Ay{0}
  \def\offset{2.5}
  \def\Bx{\Ax+5}
  \def\By{\Ay}
  \node[state,shape=circle,draw=black] (Z) at (\Ax,\Ay) {$E$};
  \node[state,shape=circle,draw=black] (Y) at (\Bx,\By) {$Y$};
  \node[state,shape=circle,draw=black] (A) at (\Bx-\offset,\By) {$M$};

  \draw [-latex] (Z) to [bend left=0] (A);
  \draw [-latex] (A) to [bend left=0] (Y);
   \draw [-latex] (Z) to [bend left=70] (Y);
\end{tikzpicture}
\caption{Causal mediation analysis.\label{fig2}}
\end{figure}

The following assumption is often assumed in causal mediation analysis \citep{robins1992identifiability,pearl2001direct}.

\begin{assumption}(Randomized treatment)\label{asm:1}
$E \indep \{Y_{e,m},M_{e}\}$.
\end{assumption}

Again, Assumption~\ref{asm:1} of randomized treatment implies that $p_{e,m_0,m_1,y_{00},y_{01}y_{10}y_{11}}$ in \eqref{eq:gen_p_def} does not depend on $e$ and we drop the index $e$ from the notation making $\mathbf p\in\Delta^{63}$. Note that these unobservable probabilities are related to the probabilities of observable events similar to before, as provided in the Supplementary Material~\ref{sec:pqm}.

Recall that the total effect can be decomposed as follows
\begin{align*}
Y_{e=1}-Y_{e=0}
=Y_{e=1,M_{e=0}} - Y_{e=0,M_{e=0}}  + Y_{e=1,M_{e=1}} - Y_{e=1,M_{e=0}},
\end{align*}
and hereinafter we abbreviate letters such as $e$ and $M_e$ in the subscript. The NDE and NIE can be written as a linear function of $\mathbf p$, and we provide the corresponding objective functions in the Supplementary Material~\ref{sec:pne}.

Hereinafter, we mainly focus on $E[Y_{1,M_0}]$ because it is the most difficult term to identify from real-world data. Note that  
\begin{align*}
E[Y_{1,M_0}]=&P(Y_{1,M_0}=1)\\
=& p_{00,0010} + p_{00,0011} + p_{00,0110} + p_{00,0111} + p_{00,1010} + p_{00,1011} + p_{00,1110}+ p_{00,1111}\\
+& p_{01,0010} + p_{01,0011} + p_{01,0110} + p_{01,0111} + p_{01,1010} + p_{01,1011} + p_{01,1110}+ p_{01,1111}\\
+& p_{10,0001} + p_{10,0011} + p_{10,0101} + p_{10,0111} + p_{10,1001} + p_{10,1011} + p_{10,1101}+ p_{10,1111}\\
+& p_{11,0001} + p_{11,0011} + p_{11,0101} + p_{11,0111} + p_{11,1001} + p_{11,1011} + p_{11,1101}+ p_{11,1111},
\end{align*}
and it arises because it is the bridge that allows this decomposition to work. We shall present how to conduct valid statistical inference based on the observed data.

\subsection{Fiducial bounds on natural effects}
Following the proposed unified framework, we consider a random vector of independent
$
V^*_{emy} \sim Beta(n_{emy}, 1), e,m,y=0,1
$ 
truncated to satisfy constraints $\mathbf{A}_{\text{obs}} \mathbf{p} \geq {\mathbf V}^*$ with the concrete form given in the Supplementary Material~\ref{sec:ne}. 

Additional assumptions are often considered to obtain informative and sharp bounds in causal mediation analysis \citep{robins2010alternative}.

\begin{assumption}\label{asm:2}
(Randomized mediator within levels of treatment)
$Y_{e,m}\indep M_e|E=e$.
\end{assumption}

The randomized mediator Assumption~\ref{asm:2} introduces additional quadratic constraints.
By Assumption~\ref{asm:2}, we have
$P(Y_{e,m}=1,M_e=1)=P(Y_{e,m}=1)P(M_e=1)$. Therefore, $\mathbf p$ must satisfy four quadratic constraints provided in the Supplementary Material~\ref{sec:add1}.
Because we have a quadratically constrained linear program (QCLP), we propose using the augmented Lagrangian adaptive barrier minimization algorithm \citep{madsen2004optimization}, which works well for nonlinear equality and inequality constraints. 

\begin{remark}
Note that under Assumptions~\ref{asm:1} and \ref{asm:2}, the fiducial bound is asymptotically equivalent to the Robins-Richardson bound \citep{robins2010alternative}. Consequently, under this setting, our approach provides valid statistical inference for the bounds.
\end{remark}

Moreover, the cross-world independence assumption assumes that
\begin{align*}
Y_{e^{*}, m} \indep M_{e},
\end{align*}
which can also be formulated as equality constraints using $\mathbf p$. We provide the details in the Supplementary Material~\ref{sec:add2}.

\subsection{Extension to population intervention indirect effect}

The population intervention indirect effect (PIIE) is another important concept from causal mediation analysis \citep{fulcher2020robust,wen2024causal,bai2025proximal}, which decomposes the population intervention effect under the non-exposure into direct and indirect components. 
The PIIE, $E[Y - Y_{E, M_0}]$, captures the indirect effect of $E$ on $Y$ and only requires an intervention of $E = 0$ on the mediator, while it does not require an intervention of $E$ on potential outcomes for $Y$. This type of indirect effect describes the extent to which an intermediate variable mediates the effect of exposure under an intervention that holds the component of exposure directly affecting the outcome at its observed level.

We essentially need to deal with $ E[Y_{E,M_0}]$, and note that
\begin{align*}
 E[Y_{E,M_0}]
=&(P(Y_{0,0}=1,M_0=0)+P(Y_{0,1}=1,M_0=1))P(E=0)\\&+
(P(Y_{1,0}=1,M_0=0)+P(Y_{1,1}=1,M_0=1))P(E=1),
\end{align*}
where the derivation is given in the Supplementary Material~\ref{sec:piieobj}.
Because 
$P(Y_{0,0}=1,M_0=0)$, 
$P(Y_{0,1}=1,M_0=1)$, 
$P(Y_{1,0}=1,M_0=0)$, and 
$P(Y_{1,1}=1,M_0=1)$
can all be written in terms of $\mathbf p$, so our framework extends to bounding the PIIE. 
The concrete form of $\mathbf{\bar A}_{\text{obs}} \mathbf{p} \geq \bar{\mathbf V}^*$ for this problem is given in the Supplementary Material~\ref{sec:piie}.

\section{Simulation studies}\label{sec:simu}
\subsection{IV bounds}

In this section, we examine the coverage and average length of 95\% fiducial confidence intervals for the lower and upper bounds, respectively. We consider the following scenario from \cite{cui2021Individualized}:
\begin{align*}
P(Y=1|U,A) &= U/16 + 1/5A + 1/15,\\
P(A=1|U,Z) &= U/16 + 2/5Z + 1/2,\\
U, Z & \sim \text{Bernoulli}(0.5).
\end{align*}
and 
a variant setting with no unmeasured confounding
\begin{align*}
P(Y=1|A) &= 1/5A + 1/15,\\
P(A=1|Z) &= 1/5Z + 1/5,\\
Z & \sim \text{Bernoulli}(0.5).
\end{align*}
We choose various sample sizes $n= 25,50,100$.
Each scenario is simulated 200 times.
The fiducial estimates are based on 1000 Monte Carlo replications.
In Tables~\ref{table:simu1}-\ref{table:simu4}, LR denotes the error rate that the true parameter is less than the lower confidence limit; UR denotes the error rate that the true parameter is greater than the upper confidence limit.
The two-sided error rate is obtained by adding the values in columns LR and UR, and the value around 5\% in aggregate indicates achieving nominal coverage. WD is the average width of the confidence interval.

We also compare the proposed method with a Bayesian approach with a Dirichlet prior \citep{loh2016noncompliance}. 
We use the recommended two priors, Dirichlet(1,1,0,0,1,1,1,1) and Dirichlet(1/2,1/2,0,0,1/4,1/4,1/4,1/4) as given in \cite{loh2016noncompliance}.
As can be seen from the tables, the proposed fiducial confidence intervals maintain the aggregate coverage and have reasonable lengths.
The proposed fiducial approach is comparable to other methods in Scenario 1 and outperforms them in Scenario 2. More importantly, we do not need to specify a particular prior.

\begin{table}[!h]
\begin{center}
\begin{tabular}{ccccccccccccc}
         & \multicolumn{3}{c}{Fiducial} & \multicolumn{3}{c}{Dirichlet prior 1}& \multicolumn{3}{c}{Dirichlet prior 2}\\
         & LR      & UR     & WD      & LR      & UR     &   WD  & LR      & UR     &   WD      \\
$n$=25  &  0.5  &  4.0  & 0.61   &  2.5   & 0.5  &  0.59  & 2.5  &  1.0 & 0.62 \\
$n$=50   & 1.5    & 3.5  &  0.46   &  3.0  &  2.0  &  0.45 &  4.0 &  2.5 &  0.46 \\
$n$=100   &  4.0   & 1.5   &  0.34  &  4.5   & 1.0   &  0.34 & 4.5 & 1.0 & 0.34 \\
\end{tabular}\\
\caption{Error rates in percent and average width of $95\%$ confidence intervals for the lower bound of Scenario 1. LR denotes the error rate that the true parameter is less than the lower confidence limit; UR denotes the error rate that the true parameter is greater than the upper confidence limit; WD is the average width of the confidence interval. Dirichlet priors 1 and 2 use Dirichlet(1,1,0,0,1,1,1,1) and Dirichlet(1/2,1/2,0,0,1/4,1/4,1/4,1/4), respectively. \label{table:simu1}}
\end{center}
\end{table}

\begin{table}[!h]
\begin{center}
\begin{tabular}{ccccccccccccc}
         & \multicolumn{3}{c}{Fiducial} & \multicolumn{3}{c}{Dirichlet prior 1}& \multicolumn{3}{c}{Dirichlet prior 2}\\
         & LR      & UR     & WD      & LR      & UR     &   WD  & LR      & UR     &   WD      \\
$n$=25  & 4.5  &  1.0  & 0.48  & 0   & 1.5  &  0.50 & 0.5 & 3.0 &0.52  \\
$n$=50   &  4.5   &  1.0  & 0.36   &  0.5   &  1.5  &  0.37 & 1.0 & 3.0 &0.37  \\
$n$=100   &  2.5   & 1.5   &  0.27  &  2.0  &  1.5 &  0.27  & 1.5 & 1.5  & 0.27 \\
\end{tabular}\\
\caption{Error rates in percent and average width of $95\%$ confidence intervals for the upper bound of Scenario 1. See Table~\ref{table:simu1} for a description of abbreviations. \label{table:simu2}}
\end{center}
\end{table}

\begin{table}[!h]
\begin{center}
\begin{tabular}{ccccccccccccc}
         & \multicolumn{3}{c}{Fiducial} & \multicolumn{3}{c}{Dirichlet prior 1}& \multicolumn{3}{c}{Dirichlet prior 2}\\
         & LR      & UR     & WD      & LR      & UR     &   WD  & LR      & UR     &   WD      \\
$n$=25  &  2.0  &  1.0  & 0.50   &  2.5   & 0  &  0.50  & 7.5  &  0.5 & 0.50 \\
$n$=50   & 1.5    & 1.5  &  0.36   &  1.5  &  0.5  &  0.36 &  4.0 &  1.0 &  0.36 \\
$n$=100   & 3.5   & 1.5   &  0.27  &  4.0   & 1.5   &  0.27 & 6.0 & 0 & 0.27 \\
\end{tabular}\\
\caption{Error rates in percent and average width of $95\%$ confidence intervals for the lower bound of Scenario 2. See Table~\ref{table:simu1} for a description of abbreviations. \label{table:simu3}}
\end{center}
\end{table}

\begin{table}[!h]
\begin{center}
\begin{tabular}{ccccccccccccc}
         & \multicolumn{3}{c}{Fiducial} & \multicolumn{3}{c}{Dirichlet prior 1}& \multicolumn{3}{c}{Dirichlet prior 2}\\
         & LR      & UR     & WD      & LR      & UR     &   WD  & LR      & UR     &   WD      \\
$n$=25  & 3.0  &  0.5  & 0.49  & 0   & 6.0  &  0.51 & 0.5 & 4.5 & 0.53  \\
$n$=50   &  1.0  &  2.5  & 0.37   &  0   &  10.5  &  0.37 & 0.5 & 6.0 &0.38  \\
$n$=100   &  2.0   & 2.5   &  0.27  &  0  &  5.0 &  0.27  & 0.5 & 4.5  & 0.27 \\
\end{tabular}\\
\caption{Error rates in percent and average width of $95\%$ confidence intervals for the upper bound of Scenario 2. See Table~\ref{table:simu1} for a description of abbreviations. \label{table:simu4}}
\end{center}
\end{table}

\subsection{Mediation analysis}
We consider the following example mediation data from \cite{demos2017language}.
\begin{align*}
P(E=1) &= 0.5,\\
P(M=1|E) &= 0.3E+0.3,\\ 
P(Y=1|M,E) &= 0.2M +0.2E+0.2.
\end{align*}
We also consider the following modified data generating process.
\begin{align*}
P(E=1) &= 0.5,\\
P(M=1|E) &= 0.2E+0.3,\\ 
P(Y=1|M,E) &= 0.3M +0.3E+0.2.
\end{align*}

We choose the sample size $n=500$, and each scenario is simulated 800 times.
The fiducial estimates are based on 1000 Monte Carlo replications. 
In Table~\ref{table:simu5}, LR denotes the error rate that the true parameter is less than the lower confidence limit; UR denotes the error rate that the true parameter is greater than the upper confidence limit.
The two-sided error rate is obtained by adding the values in columns LR and UR, and the value around 5\% in aggregate indicates achieving nominal coverage. WD is the average width of the confidence interval.

\begin{table}[!h]
\begin{center}
\begin{tabular}{cccccccccccccccc}
         & \multicolumn{3}{c}{S1} & \multicolumn{3}{c}{S2}& \multicolumn{3}{c}{S3}&
         \multicolumn{3}{c}{S4}\\
  & LR      & UR     & WD      & LR      & UR     &   WD  & LR      & UR     &   WD   & LR      & UR     &   WD    \\
      Lower bound   &  1.9  &  3.1  &  0.191 &  2.9   & 3.0  & 0.218  &  1.4 & 3.5 & 0.142 & 2.1 & 2.6 & 0.133 \\
      Upper bound   &  3.6 &  2.4  &  0.220 &   3.0   & 1.8  & 0.205 &  2.5 & 1.6 & 0.142 & 4.1 & 1.5  & 0.132 \\
\end{tabular}\\
\caption{Error rates in percent and average width of $95\%$ confidence intervals for the lower and upper bounds, respectively. LR denotes the error rate that the true parameter is less than the lower confidence limit; UR denotes the error rate that the true parameter is greater than the upper confidence limit; WD is the average width of the confidence interval.\label{table:simu5}}
\end{center}
\end{table}

\section{Real data applications}\label{sec:real}

\subsection{IV bounds}

We consider the Vitamin A example from \cite{balke1997bounds} which is a classic illustration of how causal inference can be applied to estimate the effect of an intervention subject to unmeasured confounders. 
\cite{balke1997bounds} analyzed data from a randomized experiment, where 450 villages were randomly offered oral doses of vitamin A supplementation, with 221 assigned to the control group and 229 assigned to the treatment group. 
In the study, researchers are interested in the effect of consuming Vitamin A supplementation on reducing mortality rates. 

The dataset contained 10231 individuals from villages assigned to vitamin A supplementation and 10919 untreated individuals, which defines a natural binary IV $Z$. \cite{balke1997bounds} reported bounds on the ATE, $-0.1946 \leq E(Y_1 -Y_0) \leq 0.0054$. Their conclusion is that the vitamin A supplementation, if uniformly administered, is capable of increasing the mortality rate by as much as 19.46\% and is incapable of reducing the mortality rate by more than 0.54\%.

We apply the proposed method to provide uncertainty quantification. The fiducial estimates are based on 10000 Monte Carlo replications. 
The fiducial point estimators for the bounds are the same as those reported by \cite{balke1997bounds}.
The proposed 95\% confidence interval for the lower bound is $(-0.2019,-0.1873)$  and the proposed 95\% confidence interval for the upper bound is $(0.0040, 0.0071)$.
Our estimates suggest that both bounds are statistically significant.

\subsection{Mediation analysis}
Job Search Intervention Study (JOBS II) \citep{vinokur1997mastery} is a randomized field experiment that investigates the efficacy of a job training intervention on unemployed workers, which is available in the R package \texttt{mediation}. 
In the JOBS II dataset, 1801 unemployed workers were randomly assigned to treatment and control groups.
In total, there were 899 individuals with complete information. 
The job seek measure includes two categories of high and low, where one refers to high job search self-efficacy.
A measure of depressive symptoms was also obtained and further dichotomized at the median.
Our goal is to evaluate the causal effect of the job training program on the mental health (especially through job search self-efficacy) of unemployed individuals.

We apply the proposed fiducial method to provide uncertainty quantification for the bounds. The fiducial estimates are based on 10000 Monte Carlo replications. 
The point estimators for the lower bound and upper bound are 0.093 and 0.832, respectively.
The proposed 95\% confidence interval for the lower bound of $E[Y_{e=1,M_{e=0}}]$ is $(0.012,0.174)$, and the proposed 95\% confidence interval for the upper bound is $(0.758,0.905)$. Our estimates suggest that $E[Y_{e=1,M_{e=0}}]$ is significantly bounded away from zero and one.

\section{Discussion}\label{sec:discuss}

This paper develops a unified framework for fiducial inference in partially identified models, and applies it to the IV and mediation settings. We introduce an acceptance sampling procedure to draw from the fiducial distribution, where the resulting acceptance rate serves as a diagnostic for the compatibility of the data with the identifying assumptions. We establish Bernstein–von Mises theorems for the induced fiducial distributions, providing a theoretical basis for the validity of the associated confidence intervals. We evaluate the methodology in Monte Carlo experiments and illustrate it through two real-world applications.

Several generalizations can be made following the proposed framework. First, it holds substantial promise for incorporating additional identifying assumptions, such as monotone treatment response \citep{manski1997monotone} and monotone treatment selection \citep{manski2000monotone}.
Second, the method could be adapted to a variety of models in survival analysis \citep{fleming2013counting} and proximal causal inference \citep{miao2018identifying,tchetgen2020introduction,cui2024semiparametric} under partial identification. Finally, another promising direction is to apply the approach to quantify uncertainty in program evaluation and policy learning \citep{kitagawa2018should,athey2021policy} when the underlying parameters are only partially identified \citep{cui2021Individualized,han2024optimal}.

\section{Acknowledgement}

This research was partially supported by the National Key R\&D Program of China (2024YFA1015600), the National Natural Science Foundation of China (12471266 and U23A2064), and the National Science Foundation
(DMS-2113404, 2210337, 2515303) and the United States - Israel Binational Science Foundation (BSF), Jerusalem, Israel (2024055).

\bibliographystyle{asa}
\bibliography{reference}

\begin{thebibliography}{69}
\newcommand{\enquote}[1]{``#1''}
\expandafter\ifx\csname natexlab\endcsname\relax\def\natexlab#1{#1}\fi

\bibitem[{Angrist et~al.(1996)Angrist, Imbens, and Rubin}]{angrist1996identification}
Angrist, J.~D., Imbens, G.~W., and Rubin, D.~B. (1996), \enquote{Identification of causal effects using instrumental variables,} \textit{Journal of the American statistical Association}, 91, 444--455.

\bibitem[{Athey and Wager(2021)}]{athey2021policy}
Athey, S. and Wager, S. (2021), \enquote{Policy learning with observational data,} \textit{Econometrica}, 89, 133--161.

\bibitem[{Avin et~al.(2005)Avin, Shpitser, and Pearl}]{Avin2005identifiability}
Avin, C., Shpitser, I., and Pearl, J. (2005), \enquote{Identifiability of path-specific effects,} \textit{IJCAI International Joint Conference on Artificial Intelligence}, 357--363.

\bibitem[{Bai et~al.(2025)Bai, Cui, and Sun}]{bai2025proximal}
Bai, Y., Cui, Y., and Sun, B. (2025), \enquote{Proximal Inference on Population Intervention Indirect Effect,} \textit{arXiv preprint arXiv:2504.11848}.

\bibitem[{Balke and Pearl(1994)}]{balke1994counterfactual}
Balke, A. and Pearl, J. (1994), \enquote{Counterfactual probabilities: Computational methods, bounds and applications,} in \textit{Uncertainty in artificial intelligence}, Elsevier, pp. 46--54.

\bibitem[{Balke and Pearl(1997)}]{balke1997bounds}
--- (1997), \enquote{Bounds on treatment effects from studies with imperfect compliance,} \textit{Journal of the American Statistical Association}, 92, 1171--1176.

\bibitem[{Chernozhukov et~al.(2007)Chernozhukov, Hong, and Tamer}]{chernozhukov2007estimation}
Chernozhukov, V., Hong, H., and Tamer, E. (2007), \enquote{Estimation and confidence regions for parameter sets in econometric models 1,} \textit{Econometrica}, 75, 1243--1284.

\bibitem[{Chernozhukov et~al.(2013)Chernozhukov, Lee, and Rosen}]{chernozhukov2013intersection}
Chernozhukov, V., Lee, S., and Rosen, A.~M. (2013), \enquote{Intersection bounds: Estimation and inference,} \textit{Econometrica}, 81, 667--737.

\bibitem[{Cui(2021)}]{cui2021Individualized}
Cui, Y. (2021), \enquote{Individualized {Decision}-{Making} {Under} {Partial} {Identification}: Three {Perspectives}, {Two} {Optimality} {Results}, and {One} {Paradox},} \textit{Harvard Data Science Review}, 3, 1--19.

\bibitem[{Cui and Han(2025)}]{cui2025policy}
Cui, Y. and Han, S. (2025), \enquote{Policy learning with distributional welfare,} \textit{Journal of the American Statistical Association}, 1--12.

\bibitem[{Cui and Hannig(2019)}]{cuihannig2019}
Cui, Y. and Hannig, J. (2019), \enquote{{Nonparametric generalized fiducial inference for survival functions under censoring (with discussions and rejoinder)},} \textit{Biometrika}, 106, 501--518.

\bibitem[{Cui and Hannig(2025)}]{cui2025demystifying}
--- (2025), \enquote{Demystifying inferential models and confidence curves: A fiducial perspective,} \textit{Statistical Science}, 40, 211--218.

\bibitem[{Cui et~al.(2026)Cui, Hannig, and Edlefsen}]{cui2026semiparametric}
Cui, Y., Hannig, J., and Edlefsen, P. (2026), \enquote{Semiparametric fiducial inference for Cox models,} \textit{Journal of the Royal Statistical Society Series B: Statistical Methodology}, qkag111.

\bibitem[{Cui et~al.(2024{\natexlab{a}})Cui, Hannig, and Kosorok}]{cui2024unified}
Cui, Y., Hannig, J., and Kosorok, M.~R. (2024{\natexlab{a}}), \enquote{A unified nonparametric fiducial approach to interval-censored data,} \textit{Journal of the American Statistical Association}, 119, 2230--2241.

\bibitem[{Cui et~al.(2024{\natexlab{b}})Cui, Pu, Shi, Miao, and Tchetgen~Tchetgen}]{cui2024semiparametric}
Cui, Y., Pu, H., Shi, X., Miao, W., and Tchetgen~Tchetgen, E. (2024{\natexlab{b}}), \enquote{Semiparametric proximal causal inference,} \textit{Journal of the American Statistical Association}, 119, 1348--1359.

\bibitem[{Dawid(2024)}]{dawid2024fiducial}
Dawid, P. (2024), \enquote{Fiducial inference, then and now,} in \textit{Handbook of Bayesian, Fiducial, and Frequentist Inference}, Chapman and Hall/CRC, pp. 83--105.

\bibitem[{Demos and Salas(2017)}]{demos2017language}
Demos, A. and Salas, S. (2017), \textit{A language, not a letter: Learning statistics in R}.

\bibitem[{Efron(1998)}]{Efron1998}
Efron, B. (1998), \enquote{{R.A.Fisher} in the 21st Century,} \textit{Statistical Science}, 13, 95--122.

\bibitem[{Fisher(1922)}]{Fisher1922}
Fisher, R.~A. (1922), \enquote{On the mathematical foundations of theoretical statistics,} \textit{Philosophical Transactions of the Royal Society of London. Series A}, 222, 309 -- 368.

\bibitem[{Fisher(1930)}]{Fisher1930}
--- (1930), \enquote{{Inverse probability},} \textit{Proceedings of the Cambridge Philosophical Society}, xxvi, 528--535.

\bibitem[{Fisher(1933)}]{Fisher1933}
--- (1933), \enquote{{The concepts of inverse probability and fiducial probability referring to unknown parameters},} \textit{Proceedings of the Royal Society of London Series A}, 139, 343--348.

\bibitem[{Fleming and Harrington(2013)}]{fleming2013counting}
Fleming, T.~R. and Harrington, D.~P. (2013), \textit{Counting processes and survival analysis}, vol. 625, John Wiley \& Sons.

\bibitem[{Fulcher et~al.(2020)Fulcher, Shpitser, Marealle, and Tchetgen~Tchetgen}]{fulcher2020robust}
Fulcher, I.~R., Shpitser, I., Marealle, S., and Tchetgen~Tchetgen, E.~J. (2020), \enquote{Robust inference on population indirect causal effects: the generalized front door criterion,} \textit{Journal of the Royal Statistical Society Series B: Statistical Methodology}, 82, 199--214.

\bibitem[{Han(2024)}]{han2024optimal}
Han, S. (2024), \enquote{Optimal dynamic treatment regimes and partial welfare ordering,} \textit{Journal of the American Statistical Association}, 119, 2000--2010.

\bibitem[{Hannig and Iyer(2022)}]{hannig2022testing}
Hannig, J. and Iyer, H. (2022), \enquote{Testing for calibration discrepancy of reported likelihood ratios in forensic science,} \textit{Journal of the Royal Statistical Society Series A: Statistics in Society}, 185, 267--301.

\bibitem[{Hannig et~al.(2016)Hannig, Iyer, Lai, and Lee}]{hannig2016generalized}
Hannig, J., Iyer, H., Lai, R.~C., and Lee, T.~C. (2016), \enquote{Generalized Fiducial Inference: A Review and New Results,} \textit{Journal of the American Statistical Association}, 111, 1346--1361.

\bibitem[{Hannig et~al.(2025)Hannig, Iyer, Lee, and Cui}]{hannig2025fiducial}
Hannig, J., Iyer, H., Lee, T.~C., and Cui, Y. (2025), \enquote{Fiducial Inference, On,} in \textit{International Encyclopedia of Statistical Science}, Springer, pp. 949--955.

\bibitem[{Hannig et~al.(2019)Hannig, Riman, Iyer, and Vallone}]{hannig2019reported}
Hannig, J., Riman, S., Iyer, H., and Vallone, P.~M. (2019), \enquote{Are reported likelihood ratios well calibrated?} \textit{Forensic Science International: Genetics Supplement Series}, 7, 572--574.

\bibitem[{Heckman and Vytlacil(1999)}]{heckman1999local}
Heckman, J.~J. and Vytlacil, E.~J. (1999), \enquote{Local instrumental variables and latent variable models for identifying and bounding treatment effects,} \textit{Proceedings of the National Academy of Sciences}, 96, 4730--4734.

\bibitem[{Hern{\'a}n and Robins(2023)}]{hernan2023causal}
Hern{\'a}n, M.~A. and Robins, J.~M. (2023), \textit{Causal inference: what if}, CRC PRESS.

\bibitem[{Imai et~al.(2010)Imai, Keele, and Yamamoto}]{imai2010identification}
Imai, K., Keele, L., and Yamamoto, T. (2010), \enquote{Identification, inference and sensitivity analysis for causal mediation effects,} \textit{Statistical Science}, 25, 51--71.

\bibitem[{Imbens and Angrist(1994)}]{angrist1994}
Imbens, G.~W. and Angrist, J.~D. (1994), \enquote{Identification and Estimation of Local Average Treatment Effects,} \textit{Econometrica}, 62, 467--475.

\bibitem[{Imbens and Manski(2004)}]{imbens2004confidence}
Imbens, G.~W. and Manski, C.~F. (2004), \enquote{Confidence intervals for partially identified parameters,} \textit{Econometrica}, 72, 1845--1857.

\bibitem[{Kaido et~al.(2019)Kaido, Molinari, and Stoye}]{kaido2019confidence}
Kaido, H., Molinari, F., and Stoye, J. (2019), \enquote{Confidence intervals for projections of partially identified parameters,} \textit{Econometrica}, 87, 1397--1432.

\bibitem[{Kitagawa and Tetenov(2018)}]{kitagawa2018should}
Kitagawa, T. and Tetenov, A. (2018), \enquote{Who should be treated? empirical welfare maximization methods for treatment choice,} \textit{Econometrica}, 86, 591--616.

\bibitem[{Lawrence et~al.(2024)Lawrence, Murph, Wiel, and Liu}]{lawrence2024new}
Lawrence, E.~C., Murph, A.~C., Wiel, S. A.~V., and Liu, C. (2024), \enquote{A new method for multinomial inference using Dempster-Shafer theory,} \textit{arXiv preprint arXiv:2410.05512}.

\bibitem[{Liang et~al.(2025)Liang, Kim, and Sun}]{liang2025extended}
Liang, F., Kim, S., and Sun, Y. (2025), \enquote{Extended fiducial inference: toward an automated process of statistical inference,} \textit{Journal of the Royal Statistical Society Series B: Statistical Methodology}, 87, 98--131.

\bibitem[{Liu et~al.(2019)Liu, Hannig, and Pal~Majumder}]{Liu2019}
Liu, Y., Hannig, J., and Pal~Majumder, A. (2019), \enquote{Second-Order Probability Matching Priors for the Person Parameter in Unidimensional IRT Models,} \textit{Psychometrika}, 84, 701--718.

\bibitem[{Loh and Richardson(2016)}]{loh2016noncompliance}
Loh, W.~W. and Richardson, T.~S. (2016), \textit{noncompliance: Causal Inference in the Presence of Treatment Noncompliance Under the Binary Instrumental Variable Model}, r package version 0.2.2.

\bibitem[{Madsen et~al.(2004)Madsen, Nielsen, and Tingleff}]{madsen2004optimization}
Madsen, K., Nielsen, H.~B., and Tingleff, O. (2004), \textit{Optimization with constraints}, 2nd ed.

\bibitem[{Manski(1990)}]{manski1990nonparametric}
Manski, C.~F. (1990), \enquote{Nonparametric bounds on treatment effects,} \textit{The American Economic Review}, 80, 319--323.

\bibitem[{Manski(1997)}]{manski1997monotone}
--- (1997), \enquote{Monotone treatment response,} \textit{Econometrica: Journal of the Econometric Society}, 1311--1334.

\bibitem[{Manski and Pepper(2000)}]{manski2000monotone}
Manski, C.~F. and Pepper, J.~V. (2000), \enquote{Monotone Instrumental Variables: With an Application to the Returns to Schooling,} \textit{Econometrica}, 68, 997--1010.

\bibitem[{Martin and Liu(2015)}]{martin2015inferential}
Martin, R. and Liu, C. (2015), \textit{Inferential models: Reasoning with uncertainty}, Chapman \& Hall/CRC Monographs on Statistics \& Applied Probability, CRC Press.

\bibitem[{Miao et~al.(2018)Miao, Geng, and Tchetgen~Tchetgen}]{miao2018identifying}
Miao, W., Geng, Z., and Tchetgen~Tchetgen, E.~J. (2018), \enquote{Identifying causal effects with proxy variables of an unmeasured confounder,} \textit{Biometrika}, 105, 987--993.

\bibitem[{Miles et~al.(2017)Miles, Kanki, Meloni, and Tchetgen~Tchetgen}]{miles2017partial}
Miles, C., Kanki, P., Meloni, S., and Tchetgen~Tchetgen, E. (2017), \enquote{On partial identification of the natural indirect effect,} \textit{Journal of Causal Inference}, 5, 20160004.

\bibitem[{Molchanov and Molinari(2018)}]{molchanov2018random}
Molchanov, I. and Molinari, F. (2018), \textit{Random sets in econometrics}, vol.~60, Cambridge University Press.

\bibitem[{Murph et~al.(2023)Murph, Hannig, and Williams}]{murph2023introduction}
Murph, A.~C., Hannig, J., and Williams, J.~P. (2023), \enquote{Introduction to generalized fiducial inference,} \textit{arXiv preprint arXiv:2302.14598}.

\bibitem[{Neupert et~al.(2021)Neupert, Growney, Zhu, Sorensen, Smith, and Hannig}]{neupert2021bff}
Neupert, S.~D., Growney, C.~M., Zhu, X., Sorensen, J.~K., Smith, E.~L., and Hannig, J. (2021), \enquote{BFF: bayesian, fiducial, and frequentist analysis of cognitive engagement among cognitively impaired older adults,} \textit{Entropy}, 23, 428.

\bibitem[{Pearl(2001)}]{pearl2001direct}
Pearl, J. (2001), \enquote{Direct and indirect effects,} in \textit{Proceedings of the Seventeenth Conference on Uncertainty in Artificial Intelligence}, pp. 411--420.

\bibitem[{Richardson et~al.(2015)Richardson, Hudgens, Gilbert, and Fine}]{richardson2015nonparametric}
Richardson, A., Hudgens, M.~G., Gilbert, P.~B., and Fine, J.~P. (2015), \enquote{Nonparametric bounds and sensitivity analysis of treatment effects,} \textit{Statistical Science}, 29, 596.

\bibitem[{Richardson and Robins(2010)}]{richardson2010analysis}
Richardson, T.~S. and Robins, J.~M. (2010), \enquote{Analysis of the binary instrumental variable model,} \textit{Heuristics, Probability and Causality: A Tribute to Judea Pearl}, 25, 415--444.

\bibitem[{Robins(1989)}]{robins1989analysis}
Robins, J.~M. (1989), \enquote{The analysis of randomized and non-randomized AIDS treatment trials using a new approach to causal inference in longitudinal studies,} \textit{Health service research methodology: a focus on AIDS}, 113--159.

\bibitem[{Robins and Greenland(1992)}]{robins1992identifiability}
Robins, J.~M. and Greenland, S. (1992), \enquote{Identifiability and exchangeability for direct and indirect effects,} \textit{Epidemiology}, 3, 143--155.

\bibitem[{Robins and Richardson(2010)}]{robins2010alternative}
Robins, J.~M. and Richardson, T.~S. (2010), \enquote{Alternative graphical causal models and the identification of direct effects,} \textit{Causality and psychopathology: Finding the determinants of disorders and their cures}, 84, 103--158.

\bibitem[{Schweder and Hjort(2016)}]{schweder2016confidence}
Schweder, T. and Hjort, N.~L. (2016), \textit{Confidence, likelihood, probability}, vol.~41, Cambridge University Press.

\bibitem[{Steen et~al.(2017)Steen, Loeys, Moerkerke, and Vansteelandt}]{steen2017medflex}
Steen, J., Loeys, T., Moerkerke, B., and Vansteelandt, S. (2017), \enquote{Medflex: an R package for flexible mediation analysis using natural effect models,} \textit{Journal of Statistical Software}, 76, 1--46.

\bibitem[{Stoye(2009)}]{stoye2009more}
Stoye, J. (2009), \enquote{More on confidence intervals for partially identified parameters,} \textit{Econometrica}, 77, 1299--1315.

\bibitem[{Su et~al.(2022)Su, Hannig, and Lee}]{su2022uncertainty}
Su, Y., Hannig, J., and Lee, T.~C. (2022), \enquote{Uncertainty Quantification in Graphon Estimation Using Generalized Fiducial Inference,} \textit{IEEE Transactions on Signal and Information Processing over Networks}, 8, 597--609.

\bibitem[{Swanson et~al.(2018)Swanson, Hern{\'a}n, Miller, Robins, and Richardson}]{swanson2018partial}
Swanson, S.~A., Hern{\'a}n, M.~A., Miller, M., Robins, J.~M., and Richardson, T.~S. (2018), \enquote{Partial identification of the average treatment effect using instrumental variables: review of methods for binary instruments, treatments, and outcomes,} \textit{Journal of the American Statistical Association}, 113, 933--947.

\bibitem[{Swanson et~al.(2015)Swanson, Holme, L{\o}berg, Kalager, Bretthauer, Hoff, Aas, and Hern{\'a}n}]{swanson2015bounding}
Swanson, S.~A., Holme, {\O}., L{\o}berg, M., Kalager, M., Bretthauer, M., Hoff, G., Aas, E., and Hern{\'a}n, M.~A. (2015), \enquote{Bounding the per-protocol effect in randomized trials: an application to colorectal cancer screening,} \textit{Trials}, 16, 1--11.

\bibitem[{Tchetgen~Tchetgen(2024)}]{tchetgen2024nudge}
Tchetgen~Tchetgen, E.~J. (2024), \enquote{The nudge average treatment effect,} \textit{arXiv preprint arXiv:2410.23590}.

\bibitem[{Tchetgen~Tchetgen and Phiri(2014)}]{tchetgen2014bounds}
Tchetgen~Tchetgen, E.~J. and Phiri, K. (2014), \enquote{Bounds for pure direct effect,} \textit{Epidemiology}, 25, 775--776.

\bibitem[{Tchetgen~Tchetgen and Shpitser(2012)}]{tchetgen2012semiparametric}
Tchetgen~Tchetgen, E.~J. and Shpitser, I. (2012), \enquote{Semiparametric theory for causal mediation analysis: efficiency bounds, multiple robustness, and sensitivity analysis,} \textit{Annals of statistics}, 40, 1816--1845.

\bibitem[{Tchetgen~Tchetgen et~al.(2024)Tchetgen~Tchetgen, Ying, Cui, Shi, and Miao}]{tchetgen2020introduction}
Tchetgen~Tchetgen, E.~J., Ying, A., Cui, Y., Shi, X., and Miao, W. (2024), \enquote{An introduction to proximal causal inference,} \textit{Statistical Science}, 39, 375--390.

\bibitem[{VanderWeele(2015)}]{vanderweele2015explanation}
VanderWeele, T.~J. (2015), \textit{Explanation in causal inference: methods for mediation and interaction}, Oxford University Press.

\bibitem[{Vinokur and Schul(1997)}]{vinokur1997mastery}
Vinokur, A.~D. and Schul, Y. (1997), \enquote{Mastery and inoculation against setbacks as active ingredients in the JOBS intervention for the unemployed,} \textit{Journal of Consulting and Clinical Psychology}, 65, 867--877.

\bibitem[{Wen et~al.(2024)Wen, Sarvet, and Stensrud}]{wen2024causal}
Wen, L., Sarvet, A., and Stensrud, M. (2024), \enquote{Causal effects of intervening variables in settings with unmeasured confounding,} \textit{Journal of Machine Learning Research}, 25, 1--54.

\bibitem[{Xie and Singh(2013)}]{XieSingh2013}
Xie, M. and Singh, K. (2013), \enquote{Confidence Distribution, the Frequentist Distribution Estimator of a Parameter: A Review,} \textit{International Statistical Review}, 81, 3 -- 39.

\end{thebibliography}

\linespread{1.5}\selectfont
 \numberwithin{equation}{section}

\newpage
\appendix
\begin{center}
\textbf{\Large Supplementary Material}
\end{center}

\section{Proof of Lemma~\ref{lm:Dirichlet}}
\begin{proof}
    The first assertion follows immediately by the Central Limit Theorem.

    For the second statement, let us first assume that $\mathbf N/n\to \mathbf q$ a.s. Conditionally on $\mathbf N=(n_1,\ldots,n_l)$, let $\mathbf X=(X_{0},X_{1},\ldots,X_{l})^\top$ be independent $X_{i}\sim\Gamma(n_{i},1)$ random variables, with $n_{0}=1$.  Using a coupling 
    we select a version of $\mathbf X/n\to \mathbf q^0$ whenever $\mathbf n/n\to \mathbf q$.
    A direct calculation using the characteristic function shows that 
    \[
     n^{-1/2}(\mathbf X - \mathbf N^0)\toD N(\mathbf 0,\diag(\mathbf q^0)) \mbox{ a.s., where } 
     \mathbf q_0=\begin{pmatrix}
          0\\ \mathbf  q
      \end{pmatrix}.
    \]

    Next, let $f(\mathbf X)=\mathbf X/(\mathbf 1^\top \mathbf X)\sim $ Dirichlet$(1,\mathbf n)$ and set $\mathbf V^*=f(\mathbf X)$. Using the Taylor series, we get
    \[
     f(\mathbf X/n)=f(\mathbf N^0/n)+\int_0^1 \nabla f(\mathbf N^0/n + v(\mathbf X/n-\mathbf N^0/n))dv\times (\mathbf X/n-\mathbf N^0/n)
    \]
    and 
    \begin{equation}\label{eq:Taylor1}
      \sqrt{n}\left(\mathbf V^*-\mathbf N^0/n\right)=\int_0^1 \nabla f(\mathbf N^0/n + v(\mathbf X/n-\mathbf N^0/n))dv\times n^{-1/2}(\mathbf X-\mathbf N^0).
    \end{equation}
    The integral on the right hand side of \eqref{eq:Taylor1} converges to $\mathbf I-\mathbf q^0\mathbf 1^\top$ a.s. and consequently
    $\sqrt{n}\left(\mathbf V^*-\mathbf N^0/n\right)\toD N(\mathbf 0,\Sigma^0)$ a.s.

    Finally, if $\mathbf N/n\toP \mathbf q$ only, the theorem follows by use of the subsequence criterion for convergence in probability and the Prokhorov metric between the distributions of $\sqrt{n}\left(\mathbf V^*-\mathbf N^0/n\right)$ and $N(\mathbf 0,\Sigma^0)$.
\end{proof}

\section{Proof of Theorem~\ref{thm:acceptance}}

\begin{proof}
    Lemma~\ref{lm:Dirichlet} and Equation~\eqref{eq:Sz} imply that $\sqrt{n_c}\operatorname{diam} \mathcal S_c\toP 0$ as $n_c\to\infty$. Moreover, for any open ball $\mathcal B$ containing $\mathbf q$,  
    $P^*(\mathcal S_0\times\mathcal S_1 \subset \mathcal B)\toP 1$. Statements 1 and 2 follow immediately.

  Next, notice that $\mathcal F$ is convex. If $\mathbf q\in \partial \mathcal F$ then there is a supporting hyperplane and a corresponding normal vector $\mathbf v$ pointing out of $\mathcal F$. Notice that 
  \begin{equation}\label{eq:acceptaceComp}
  P^*((\mathcal S_0\times\mathcal S_1)\cap\mathcal F=\emptyset)\geq
  P^*(\min \mathbf v^\top (\mathcal S_0\times\mathcal S_1)>\mathbf v^\top \mathbf q).
  \end{equation}
   Lemma~\ref{lm:Dirichlet} implies that
  \begin{equation}\label{eq:normalMult2}
    \sqrt{n_0+n_1}((\mathbf N_0/n_0,\mathbf N_1/n_1)-\mathbf q)\toD \mathbf Z\sim N(\mathbf 0,\Sigma),
    \end{equation}
where
\[
    \Sigma = \begin{pmatrix}
               \lambda_0^{-1}( \diag(\mathbf q_0)-\mathbf q_0\mathbf q_0^\top )& 0\\
                0 & \lambda_1^{-1}(\diag(\mathbf q_1)-\mathbf q_1\mathbf q_1^\top)
    \end{pmatrix}.
  \]
  Moreover, $\mathbf v$ can be chosen so that  $\mathbf v^\top\mathbf  Z$ is a non-degenerate, mean zero, Gaussian random variable. Denote $F(s)$ its distribution function.
  
  Skorokhod representation theorem allows us to select a sequence of multinomials for which the convergence in \eqref{eq:normalMult2} is a.s. For such a sequence 
  \[
    \sqrt{n_0+n_1}((\mathcal S_0\times\mathcal S_1) -(\mathbf N_0/n_0,\mathbf N_1/n_1))\toD N(\mathbf 0,\Sigma) \mbox{ a.s.}
  \]
  Thus by subtracting $\mathbf v^\top (\mathbf N_0/n_0,\mathbf N_1/n_1)$ and multiplying by $\sqrt{n_0+n_1}$ on both sides one gets
  \[
    P^*(\min \mathbf v^\top (\mathcal S_0\times\mathcal S_1)>\mathbf v^\top \mathbf q)\to F(-\mathbf v^\top\mathbf Z)\sim U(0,1)\mbox{ a.s.}
  \]
  The assertion follows. 
\end{proof}

\section{Proof of Theorems~\ref{thm:BvM2} and \ref{thm:BvM2A}}

\begin{proof}
    Theorem~\ref{thm:acceptance} implies that the fiducial probability that $G^*$ exists, i.e., $\mathbf V^*$ is such that the optimization problem is feasible, goes to one in probability. Consequently, the result for both theorems is a direct application of the delta method and the asymptotic properties of multinomial and Dirichlet distributions. The calculations are straightforward and are omitted to save space. 
\end{proof}

\section{Proof of Proposition~\ref{prop:qp}}
\begin{proof}
We first prove the first equation in \eqref{eq:q2p}. By definition, we know that
\begin{align*}
& p_{00,01} +p_{00,00} + p_{01,01} + p_{01,00}  \\= & P(A_0=0,A_1=0,Y_0=0,Y_1=1) + P(A_0=0,A_1=0,Y_0=0,Y_1=0)\\ 
&  + P(A_0=0,A_1=1,Y_0=0,Y_1=1)+ P(A_0=0,A_1=1,Y_0=0,Y_1=0)\\
= & P(A_0=0,Y_0=0,Y_1=1) + P(A_0=0,Y_0=0,Y_1=0)\\
= & P(A_0=0,Y_0=0)\\
= & P(A_0=0,Y_0=0|Z=0)~~~\text{(by~Assumption~\ref{asm:ivindep})}\\
= & P(A=0,Y=0|Z=0)
=q_{000}.
\end{align*}
The other equalities are proved analogously.
\end{proof}

\section{Relaxing IV independence Assumption~\ref{asm:ivindep}}\label{sec:32numbers}

Relaxing Assumption~\ref{asm:ivindep} leads to a modification of \eqref{eq:causalDSM}
\begin{equation}\label{eq:causalDSM32}
    \begin{aligned}
V_{000}^* &\leq p_{0,00,00} + p_{0,00,01} +  p_{0,01,00} + p_{0,01,01}\\
V_{001}^* &\leq p_{0,00,10} + p_{0,00,11} + p_{0,01,10} + p_{0,01,11} \\
V_{010}^* &\leq p_{0,10,00} + p_{0,10,10} + p_{0,11,00} + p_{0,11,10} \\
V_{011}^* &\leq p_{0,10,01} + p_{0,10,11} + p_{0,11,01} + p_{0,11,11} \\
V_{100}^* &\leq p_{1,00,00} + p_{1,00,01} + p_{1,10,00} + p_{1,10,01} \\
V_{101}^* &\leq p_{1,00,10} + p_{1,00,11} + p_{1,10,10} + p_{1,10,11} \\
V_{110}^* &\leq p_{1,01,00} + p_{1,01,10} + p_{1,11,00} + p_{1,11,10} \\
V_{111}^* &\leq p_{1,01,01} + p_{1,01,11} + p_{1,11,01} + p_{1,11,11},       
    \end{aligned}
\end{equation}
where $p_{z,a_0a_1,y_0y_1}=P(A_0=a_0,A_1=a_1,Y_0=y_0,Y_1=y_1|Z=z)$, $V^*_{0ay} \sim$ Beta$(n_{0ay}, 1)$, and $V^*_{1ay} \sim$ Beta$(n_{1ay}, 1)$. 

\section{Detailed constraints and objective functions}
\subsection{Equality constraints related $\mathbf p$ to $\mathbf q$ in mediation analysis}\label{sec:pqm}

\begin{align*}
q_{000} &=(p_{00,0000} + p_{00,0001} +p_{00,0010} + p_{00,0011} +p_{00,0100} + p_{00,0101} +p_{00,0110} + p_{00,0111}
\\&+  p_{01,0000} + p_{01,0001} +p_{01,0010} + p_{01,0011} +p_{01,0100} + p_{01,0101} +p_{01,0110} + p_{01,0111}),\\
q_{001} &=(p_{00,1000} + p_{00,1001} +p_{00,1010} + p_{00,1011} +p_{00,1100} + p_{00,1101} +p_{00,1110} + p_{00,1111}
\\&+  p_{01,1000} + p_{01,1001} +p_{01,1010} + p_{01,1011} +p_{01,1100} + p_{01,1101} +p_{01,1110} + p_{01,1111}),\\
q_{010} &= (p_{11,0000} + p_{11,0001} +p_{11,0010} + p_{11,0011} +p_{11,1000} + p_{11,1001} +p_{11,1010} + p_{11,1011}
\\&+  p_{10,0000} + p_{10,0001} +p_{10,0010} + p_{10,0011} +p_{10,1000} + p_{10,1001} +p_{10,1010} + p_{10,1011}), \\
q_{011} &= (p_{11,0100} + p_{11,0101} +p_{11,0110} + p_{11,0111} +p_{11,1100} + p_{11,1101} +p_{11,1110} + p_{11,1111}
\\&+  p_{10,0100} + p_{10,0101} +p_{10,0110} + p_{10,0111} +p_{10,1100} + p_{10,1101} +p_{10,1110} + p_{10,1111}), \\
q_{100} &= (p_{00,0000} + p_{00,0001} +p_{00,0100} + p_{00,0101} +p_{00,1000} + p_{00,1001} +p_{00,1100} + p_{00,1101}
\\&+  p_{10,0000} + p_{10,0001} +p_{10,0100} + p_{10,0101} +p_{10,1000} + p_{10,1001} +p_{10,1100} + p_{10,1101}), \\
q_{101} &= (p_{00,0010} + p_{00,0011} +p_{00,0110} + p_{00,0111} +p_{00,1010} + p_{00,1011} +p_{00,1110} + p_{00,1111}
\\&+  p_{10,0010} + p_{10,0011} +p_{10,0110} + p_{10,0111} +p_{10,1010} + p_{10,1011} +p_{10,1110} + p_{10,1111}),\\
q_{110} &= (p_{11,0000} + p_{11,0010} +p_{11,0100} + p_{11,0110} +p_{11,1000} + p_{11,1010} +p_{11,1100} + p_{11,1110}
\\&+  p_{01,0000} + p_{01,0010} +p_{01,0100} + p_{01,0110} +p_{01,1000} + p_{01,1010} +p_{01,1100} + p_{01,1110}), \\
q_{111} &= (p_{11,0001} + p_{11,0011} +p_{11,0101} + p_{11,0111} +p_{11,1001} + p_{11,1011} +p_{11,1101} + p_{11,1111}
\\&+  p_{01,0001} + p_{01,0011} +p_{01,0101} + p_{01,0111} +p_{01,1001} + p_{01,1011} +p_{01,1101} + p_{01,1111}).
\end{align*}

\subsection{Objective functions for NDE and NIE}\label{sec:pne}

The NDE and NIE can be written as
\begin{align*}
& E[Y_{1,M_{0}} - Y_{0,M_{0}}] \\
= & p_{00,0010} - p_{00,1000}
+ p_{00,0110} - p_{00,1100}
+ p_{00,0011} - p_{00,1001}
+ p_{00,0111} - p_{00,1101}\\
+ & p_{11,0001} - p_{11,0100}
+ p_{11,0011} - p_{11,0110}
+ p_{11,1001} - p_{11,1100}
+ p_{11,1011} - p_{11,1110}\\
+ & p_{01,0010} - p_{01,1000}
+ p_{01,0110} - p_{01,1100}
+ p_{01,0011} - p_{01,1001}
+ p_{01,0111} - p_{01,1101}\\
+ & p_{10,0001} - p_{10,0100}
+ p_{10,0011} - p_{10,0110}
+ p_{10,1001} - p_{10,1100}
+ p_{10,1011} - p_{10,1110},\\
& E[Y_{1,M_{1}} - Y_{1,M_{0}}] \\
= & p_{01,0001} - p_{01,0010}
+ p_{01,0101} - p_{01,0110}
+ p_{01,1001} - p_{01,1010}
+ p_{01,1101} - p_{01,1110}\\
+ & p_{10,0010} - p_{10,0001}
+ p_{10,0110} - p_{10,0101}
+ p_{10,1010} - p_{10,1001}
+ p_{10,1110} - p_{10,1101},
\end{align*}
respectively. 

\subsection{Constraints for natural effects}\label{sec:ne}

\begin{align*}
V_{000}^* &\leq (p_{00,0000} + p_{00,0001} +p_{00,0010} + p_{00,0011} +p_{00,0100} + p_{00,0101} +p_{00,0110} + p_{00,0111}
\\&+  p_{01,0000} + p_{01,0001} +p_{01,0010} + p_{01,0011} +p_{01,0100} + p_{01,0101} +p_{01,0110} + p_{01,0111}), \\
V_{001}^* &\leq (p_{00,1000} + p_{00,1001} +p_{00,1010} + p_{00,1011} +p_{00,1100} + p_{00,1101} +p_{00,1110} + p_{00,1111}
\\&+  p_{01,1000} + p_{01,1001} +p_{01,1010} + p_{01,1011} +p_{01,1100} + p_{01,1101} +p_{01,1110} + p_{01,1111}), \\
V_{010}^* &\leq (p_{11,0000} + p_{11,0001} +p_{11,0010} + p_{11,0011} +p_{11,1000} + p_{11,1001} +p_{11,1010} + p_{11,1011}
\\&+  p_{10,0000} + p_{10,0001} +p_{10,0010} + p_{10,0011} +p_{10,1000} + p_{10,1001} +p_{10,1010} + p_{10,1011}), \\
V_{011}^* &\leq (p_{11,0100} + p_{11,0101} +p_{11,0110} + p_{11,0111} +p_{11,1100} + p_{11,1101} +p_{11,1110} + p_{11,1111}
\\&+  p_{10,0100} + p_{10,0101} +p_{10,0110} + p_{10,0111} +p_{10,1100} + p_{10,1101} +p_{10,1110} + p_{10,1111}), \\
V_{100}^* &\leq (p_{00,0000} + p_{00,0001} +p_{00,0100} + p_{00,0101} +p_{00,1000} + p_{00,1001} +p_{00,1100} + p_{00,1101}
\\&+  p_{10,0000} + p_{10,0001} +p_{10,0100} + p_{10,0101} +p_{10,1000} + p_{10,1001} +p_{10,1100} + p_{10,1101}),\\
V_{101}^* &\leq (p_{00,0010} + p_{00,0011} +p_{00,0110} + p_{00,0111} +p_{00,1010} + p_{00,1011} +p_{00,1110} + p_{00,1111}
\\&+  p_{10,0010} + p_{10,0011} +p_{10,0110} + p_{10,0111} +p_{10,1010} + p_{10,1011} +p_{10,1110} + p_{10,1111}),\\
V_{110}^* &\leq (p_{11,0000} + p_{11,0010} +p_{11,0100} + p_{11,0110} +p_{11,1000} + p_{11,1010} +p_{11,1100} + p_{11,1110}
\\&+  p_{01,0000} + p_{01,0010} +p_{01,0100} + p_{01,0110} +p_{01,1000} + p_{01,1010} +p_{01,1100} + p_{01,1110}),\\
V_{111}^* &\leq (p_{11,0001} + p_{11,0011} +p_{11,0101} + p_{11,0111} +p_{11,1001} + p_{11,1011} +p_{11,1101} + p_{11,1111}
\\&+  p_{01,0001} + p_{01,0011} +p_{01,0101} + p_{01,0111} +p_{01,1001} + p_{01,1011} +p_{01,1101} + p_{01,1111}).
\end{align*}

\subsection{Additional four constraints implied by Assumption~\ref{asm:2}}\label{sec:add1}
The constraints for $(e=0,m=0)$,
$(e=0,m=1)$,
$(e=1,m=0)$, and
$(e=1,m=1)$ are given below, respectively.
\begin{align*}
& (p_{11,1000}+p_{11,1001}+p_{11,1010}+p_{11,1011}+p_{11,1100}+p_{11,1101}+p_{11,1110}+p_{11,1111}\\
&+ p_{10,1000}+p_{10,1001}+p_{10,1010}+p_{10,1011}+p_{10,1100}+p_{10,1101}+p_{10,1110}+p_{10,1111})\\
&=
(p_{11,1000}+p_{11,1001}+p_{11,1010}+p_{11,1011}+p_{11,1100}+p_{11,1101}+p_{11,1110}+p_{11,1111}\\
& + p_{10,1000}+p_{10,1001}+p_{10,1010}+p_{10,1011}+p_{10,1100}+p_{10,1101}+p_{10,1110}+p_{10,1111}\\
& + p_{00,1000}+p_{00,1001}+p_{00,1010}+p_{00,1011}+p_{00,1100}+p_{00,1101}+p_{00,1110}+p_{00,1111}\\
&+ p_{01,1000}+p_{01,1001}+p_{01,1010}+p_{01,1011}+p_{01,1100}+p_{01,1101}+p_{01,1110}+p_{01,1111})\\&\times (p_{11,0000}+p_{11,0001}+p_{11,0010}+p_{11,0011}+p_{11,0100}+p_{11,0101}+p_{11,0110}+p_{11,0111}\\&+p_{11,1000}+p_{11,1001}+p_{11,1010}+p_{11,1011}+p_{11,1100}+p_{11,1101}+p_{11,1110}+p_{11,1111}\\&+p_{10,0000}+p_{10,0001}+p_{10,0010}+p_{10,0011}+p_{10,0100}+p_{10,0101}+p_{10,0110}+p_{10,0111}\\&+p_{10,1000}+p_{10,1001}+p_{10,1010}+p_{10,1011}+p_{10,1100}+p_{10,1101}+p_{10,1110}+p_{10,1111}),\\
& (p_{11,0100}+p_{11,0101}+p_{11,0110}+p_{11,0111}+p_{11,1100}+p_{11,1101}+p_{11,1110}+p_{11,1111}\\
&+ p_{10,0100}+p_{10,0101}+p_{10,0110}+p_{10,0111}+p_{10,1100}+p_{10,1101}+p_{10,1110}+p_{10,1111})\\
&=
(p_{11,0100}+p_{11,0101}+p_{11,0110}+p_{11,0111}+p_{11,1100}+p_{11,1101}+p_{11,1110}+p_{11,1111}\\
&+ p_{10,0100}+p_{10,0101}+p_{10,0110}+p_{10,0111}+p_{10,1100}+p_{10,1101}+p_{10,1110}+p_{10,1111}\\
& + p_{00,0100}+p_{00,0101}+p_{00,0110}+p_{00,0111}+p_{00,1100}+p_{00,1101}+p_{00,1110}+p_{00,1111}\\
&+ p_{01,0100}+p_{01,0101}+p_{01,0110}+p_{01,0111}+p_{01,1100}+p_{01,1101}+p_{01,1110}+p_{01,1111})\\&\times (p_{11,0000}+p_{11,0001}+p_{11,0010}+p_{11,0011}+p_{11,0100}+p_{11,0101}+p_{11,0110}+p_{11,0111}\\&+p_{11,1000}+p_{11,1001}+p_{11,1010}+p_{11,1011}+p_{11,1100}+p_{11,1101}+p_{11,1110}+p_{11,1111}\\&+p_{10,0000}+p_{10,0001}+p_{10,0010}+p_{10,0011}+p_{10,0100}+p_{10,0101}+p_{10,0110}+p_{10,0111}\\&+p_{10,1000}+p_{10,1001}+p_{10,1010}+p_{10,1011}+p_{10,1100}+p_{10,1101}+p_{10,1110}+p_{10,1111}),
\end{align*}
\begin{align*}
& (p_{11,0010}+p_{11,0011}+p_{11,0110}+p_{11,0111}+p_{11,1010}+p_{11,1011}+p_{11,1110}+p_{11,1111}\\
&+ p_{01,0010}+p_{01,0011}+p_{01,0110}+p_{01,0111}+p_{01,1010}+p_{01,1011}+p_{01,1110}+p_{01,1111})\\
&=
(p_{11,0010}+p_{11,0011}+p_{11,0110}+p_{11,0111}+p_{11,1010}+p_{11,1011}+p_{11,1110}+p_{11,1111}\\
&+ p_{01,0010}+p_{01,0011}+p_{01,0110}+p_{01,0111}+p_{01,1010}+p_{01,1011}+p_{01,1110}+p_{01,1111}\\
& + p_{00,0010}+p_{00,0011}+p_{00,0110}+p_{00,0111}+p_{00,1010}+p_{00,1011}+p_{00,1110}+p_{00,1111}\\
&+ p_{10,0010}+p_{10,0011}+p_{10,0110}+p_{10,0111}+p_{10,1010}+p_{10,1011}+p_{10,1110}+p_{10,1111})\\&\times (p_{11,0000}+p_{11,0001}+p_{11,0010}+p_{11,0011}+p_{11,0100}+p_{11,0101}+p_{11,0110}+p_{11,0111}\\&+p_{11,1000}+p_{11,1001}+p_{11,1010}+p_{11,1011}+p_{11,1100}+p_{11,1101}+p_{11,1110}+p_{11,1111}\\&+p_{01,0000}+p_{01,0001}+p_{01,0010}+p_{01,0011}+p_{01,0100}+p_{01,0101}+p_{01,0110}+p_{01,0111}\\&+p_{01,1000}+p_{01,1001}+p_{01,1010}+p_{01,1011}+p_{01,1100}+p_{01,1101}+p_{01,1110}+p_{01,1111}),\\
& (p_{11,0001}+p_{11,0011}+p_{11,0101}+p_{11,0111}+p_{11,1001}+p_{11,1011}+p_{11,1101}+p_{11,1111}\\
&+ p_{01,0001}+p_{01,0011}+p_{01,0101}+p_{01,0111}+p_{01,1001}+p_{01,1011}+p_{01,1101}+p_{01,1111})\\
&=
(p_{11,0001}+p_{11,0011}+p_{11,0101}+p_{11,0111}+p_{11,1001}+p_{11,1011}+p_{11,1101}+p_{11,1111}\\
&+ p_{01,0001}+p_{01,0011}+p_{01,0101}+p_{01,0111}+p_{01,1001}+p_{01,1011}+p_{01,1101}+p_{01,1111}\\
& + p_{00,0001}+p_{00,0011}+p_{00,0101}+p_{00,0111}+p_{00,1001}+p_{00,1011}+p_{00,1101}+p_{00,1111}\\
&+ p_{10,0001}+p_{10,0011}+p_{10,0101}+p_{10,0111}+p_{10,1001}+p_{10,1011}+p_{10,1101}+p_{10,1111})\\&\times (p_{11,0000}+p_{11,0001}+p_{11,0010}+p_{11,0011}+p_{11,0100}+p_{11,0101}+p_{11,0110}+p_{11,0111}\\&+p_{11,1000}+p_{11,1001}+p_{11,1010}+p_{11,1011}+p_{11,1100}+p_{11,1101}+p_{11,1110}+p_{11,1111}\\&+p_{01,0000}+p_{01,0001}+p_{01,0010}+p_{01,0011}+p_{01,0100}+p_{01,0101}+p_{01,0110}+p_{01,0111}\\&+p_{01,1000}+p_{01,1001}+p_{01,1010}+p_{01,1011}+p_{01,1100}+p_{01,1101}+p_{01,1110}+p_{01,1111}).\\
\end{align*}

\subsection{Additional four constraints implied by the cross-world independence}\label{sec:add2}

\begin{align*}
& (p_{11,0010}+p_{11,0011}+p_{11,0110}+p_{11,0111}+p_{11,1010}+p_{11,1011}+p_{11,1110}+p_{11,1111}\\
&+ p_{10,0010}+p_{10,0011}+p_{10,0110}+p_{10,0111}+p_{10,1010}+p_{10,1011}+p_{10,1110}+p_{10,1111})\\
&=
(p_{11,0010}+p_{11,0011}+p_{11,0110}+p_{11,0111}+p_{11,1010}+p_{11,1011}+p_{11,1110}+p_{11,1111}\\
&+ p_{01,0010}+p_{01,0011}+p_{01,0110}+p_{01,0111}+p_{01,1010}+p_{01,1011}+p_{01,1110}+p_{01,1111}\\
& + p_{00,0010}+p_{00,0011}+p_{00,0110}+p_{00,0111}+p_{00,1010}+p_{00,1011}+p_{00,1110}+p_{00,1111}\\
&+ p_{10,0010}+p_{10,0011}+p_{10,0110}+p_{10,0111}+p_{10,1010}+p_{10,1011}+p_{10,1110}+p_{10,1111})\\&\times (p_{11,0000}+p_{11,0001}+p_{11,0010}+p_{11,0011}+p_{11,0100}+p_{11,0101}+p_{11,0110}+p_{11,0111}\\&+p_{11,1000}+p_{11,1001}+p_{11,1010}+p_{11,1011}+p_{11,1100}+p_{11,1101}+p_{11,1110}+p_{11,1111}\\&+p_{10,0000}+p_{10,0001}+p_{10,0010}+p_{10,0011}+p_{10,0100}+p_{10,0101}+p_{10,0110}+p_{10,0111}\\&+p_{10,1000}+p_{10,1001}+p_{10,1010}+p_{10,1011}+p_{10,1100}+p_{10,1101}+p_{10,1110}+p_{10,1111}),\\
& (p_{11,0001}+p_{11,0011}+p_{11,0101}+p_{11,0111}+p_{11,1001}+p_{11,1011}+p_{11,1101}+p_{11,1111}\\
&+ p_{10,0001}+p_{10,0011}+p_{10,0101}+p_{10,0111}+p_{10,1001}+p_{10,1011}+p_{10,1101}+p_{10,1111})\\
&=
(p_{11,0001}+p_{11,0011}+p_{11,0101}+p_{11,0111}+p_{11,1001}+p_{11,1011}+p_{11,1101}+p_{11,1111}\\
&+ p_{01,0001}+p_{01,0011}+p_{01,0101}+p_{01,0111}+p_{01,1001}+p_{01,1011}+p_{01,1101}+p_{01,1111}\\
& + p_{00,0001}+p_{00,0011}+p_{00,0101}+p_{00,0111}+p_{00,1001}+p_{00,1011}+p_{00,1101}+p_{00,1111}\\
&+ p_{10,0001}+p_{10,0011}+p_{10,0101}+p_{10,0111}+p_{10,1001}+p_{10,1011}+p_{10,1101}+p_{10,1111})\\&\times (p_{11,0000}+p_{11,0001}+p_{11,0010}+p_{11,0011}+p_{11,0100}+p_{11,0101}+p_{11,0110}+p_{11,0111}\\&+p_{11,1000}+p_{11,1001}+p_{11,1010}+p_{11,1011}+p_{11,1100}+p_{11,1101}+p_{11,1110}+p_{11,1111}\\&+p_{10,0000}+p_{10,0001}+p_{10,0010}+p_{10,0011}+p_{10,0100}+p_{10,0101}+p_{10,0110}+p_{10,0111}\\&+p_{10,1000}+p_{10,1001}+p_{10,1010}+p_{10,1011}+p_{10,1100}+p_{10,1101}+p_{10,1110}+p_{10,1111}),
\end{align*}
\begin{align*}
& (p_{11,1000}+p_{11,1001}+p_{11,1010}+p_{11,1011}+p_{11,1100}+p_{11,1101}+p_{11,1110}+p_{11,1111}\\
&+ p_{01,1000}+p_{01,1001}+p_{01,1010}+p_{01,1011}+p_{01,1100}+p_{01,1101}+p_{01,1110}+p_{01,1111})\\
&=
(p_{11,1000}+p_{11,1001}+p_{11,1010}+p_{11,1011}+p_{11,1100}+p_{11,1101}+p_{11,1110}+p_{11,1111}\\
& + p_{10,1000}+p_{10,1001}+p_{10,1010}+p_{10,1011}+p_{10,1100}+p_{10,1101}+p_{10,1110}+p_{10,1111}\\
& + p_{00,1000}+p_{00,1001}+p_{00,1010}+p_{00,1011}+p_{00,1100}+p_{00,1101}+p_{00,1110}+p_{00,1111}\\
&+ p_{01,1000}+p_{01,1001}+p_{01,1010}+p_{01,1011}+p_{01,1100}+p_{01,1101}+p_{01,1110}+p_{01,1111})\\&\times (p_{11,0000}+p_{11,0001}+p_{11,0010}+p_{11,0011}+p_{11,0100}+p_{11,0101}+p_{11,0110}+p_{11,0111}\\&+p_{11,1000}+p_{11,1001}+p_{11,1010}+p_{11,1011}+p_{11,1100}+p_{11,1101}+p_{11,1110}+p_{11,1111}\\&+p_{01,0000}+p_{01,0001}+p_{01,0010}+p_{01,0011}+p_{01,0100}+p_{01,0101}+p_{01,0110}+p_{01,0111}\\&+p_{01,1000}+p_{01,1001}+p_{01,1010}+p_{01,1011}+p_{01,1100}+p_{01,1101}+p_{01,1110}+p_{01,1111}),\\
& (p_{11,0100}+p_{11,0101}+p_{11,0110}+p_{11,0111}+p_{11,1100}+p_{11,1101}+p_{11,1110}+p_{11,1111}\\
&+ p_{01,0100}+p_{01,0101}+p_{01,0110}+p_{01,0111}+p_{01,1100}+p_{01,1101}+p_{01,1110}+p_{01,1111})\\
&=
(p_{11,0100}+p_{11,0101}+p_{11,0110}+p_{11,0111}+p_{11,1100}+p_{11,1101}+p_{11,1110}+p_{11,1111}\\
&+ p_{10,0100}+p_{10,0101}+p_{10,0110}+p_{10,0111}+p_{10,1100}+p_{10,1101}+p_{10,1110}+p_{10,1111}\\
& + p_{00,0100}+p_{00,0101}+p_{00,0110}+p_{00,0111}+p_{00,1100}+p_{00,1101}+p_{00,1110}+p_{00,1111}\\
&+ p_{01,0100}+p_{01,0101}+p_{01,0110}+p_{01,0111}+p_{01,1100}+p_{01,1101}+p_{01,1110}+p_{01,1111})\\&\times (p_{11,0000}+p_{11,0001}+p_{11,0010}+p_{11,0011}+p_{11,0100}+p_{11,0101}+p_{11,0110}+p_{11,0111}\\&+p_{11,1000}+p_{11,1001}+p_{11,1010}+p_{11,1011}+p_{11,1100}+p_{11,1101}+p_{11,1110}+p_{11,1111}\\&+p_{01,0000}+p_{01,0001}+p_{01,0010}+p_{01,0011}+p_{01,0100}+p_{01,0101}+p_{01,0110}+p_{01,0111}\\&+p_{01,1000}+p_{01,1001}+p_{01,1010}+p_{01,1011}+p_{01,1100}+p_{01,1101}+p_{01,1110}+p_{01,1111}).\\
\end{align*}

\subsection{Objective functions for PIIE}\label{sec:piieobj}
Note that 
\begin{align*}
 E[Y_{E,M_0}]
=&E[E[E[Y_{E,M_0}\mid E,M_0]\mid E]]\\ 
=&E[E[Y_{0,M_0}\mid E=0,M_0]\mid E=0]P(E=0)+
E[E[Y_{1,M_0}\mid E=1,M_0]\mid E=1]P(E=1)\\
=&P(Y_{0,0}=1,M_0=0\mid E=0)P(E=0)+P(Y_{0,1}=1,M_0=1\mid E=0)P(E=0)\\&+
P(Y_{1,0}=1,M_0=0\mid E=1)P(E=1)+
P(Y_{1,1}=1,M_0=1\mid E=1)P(E=1)\\
=&(P(Y_{0,0}=1,M_0=0)+P(Y_{0,1}=1,M_0=1))P(E=0)\\&+
(P(Y_{1,0}=1,M_0=0)+P(Y_{1,1}=1,M_0=1))P(E=1).
\end{align*}

\subsection{Constraints for PIIE}\label{sec:piie}

\begin{align*}
\bar V_{000}^* &\leq (p_{00,0000} + p_{00,0001} +p_{00,0010} + p_{00,0011} +p_{00,0100} + p_{00,0101} +p_{00,0110} + p_{00,0111}
\\&+  p_{01,0000} + p_{01,0001} +p_{01,0010} + p_{01,0011} +p_{01,0100} + p_{01,0101} +p_{01,0110} + p_{01,0111}) (1-\rho),\\
\bar V_{001}^* &\leq (p_{00,1000} + p_{00,1001} +p_{00,1010} + p_{00,1011} +p_{00,1100} + p_{00,1101} +p_{00,1110} + p_{00,1111}
\\&+  p_{01,1000} + p_{01,1001} +p_{01,1010} + p_{01,1011} +p_{01,1100} + p_{01,1101} +p_{01,1110} + p_{01,1111}) (1-\rho),\\
\bar V_{010}^* &\leq (p_{11,0000} + p_{11,0001} +p_{11,0010} + p_{11,0011} +p_{11,1000} + p_{11,1001} +p_{11,1010} + p_{11,1011}
\\&+  p_{10,0000} + p_{10,0001} +p_{10,0010} + p_{10,0011} +p_{10,1000} + p_{10,1001} +p_{10,1010} + p_{10,1011})(1-\rho), \\
\bar V_{011}^* &\leq (p_{11,0100} + p_{11,0101} +p_{11,0110} + p_{11,0111} +p_{11,1100} + p_{11,1101} +p_{11,1110} + p_{11,1111}
\\&+  p_{10,0100} + p_{10,0101} +p_{10,0110} + p_{10,0111} +p_{10,1100} + p_{10,1101} +p_{10,1110} + p_{10,1111}) (1-\rho),\\
\bar V_{100}^* &\leq (p_{00,0000} + p_{00,0001} +p_{00,0100} + p_{00,0101} +p_{00,1000} + p_{00,1001} +p_{00,1100} + p_{00,1101}
\\&+  p_{10,0000} + p_{10,0001} +p_{10,0100} + p_{10,0101} +p_{10,1000} + p_{10,1001} +p_{10,1100} + p_{10,1101})\rho,\\
\bar V_{101}^* &\leq (p_{00,0010} + p_{00,0011} +p_{00,0110} + p_{00,0111} +p_{00,1010} + p_{00,1011} +p_{00,1110} + p_{00,1111}
\\&+  p_{10,0010} + p_{10,0011} +p_{10,0110} + p_{10,0111} +p_{10,1010} + p_{10,1011} +p_{10,1110} + p_{10,1111})\rho,\\
\bar V_{110}^* &\leq (p_{11,0000} + p_{11,0010} +p_{11,0100} + p_{11,0110} +p_{11,1000} + p_{11,1010} +p_{11,1100} + p_{11,1110}
\\&+  p_{01,0000} + p_{01,0010} +p_{01,0100} + p_{01,0110} +p_{01,1000} + p_{01,1010} +p_{01,1100} + p_{01,1110})\rho,\\
\bar V_{111}^* &\leq (p_{11,0001} + p_{11,0011} +p_{11,0101} + p_{11,0111} +p_{11,1001} + p_{11,1011} +p_{11,1101} + p_{11,1111}
\\&+  p_{01,0001} + p_{01,0011} +p_{01,0101} + p_{01,0111} +p_{01,1001} + p_{01,1011} +p_{01,1101} + p_{01,1111})\rho.
\end{align*}

\section{Other causal estimands}\label{sec:estimand}

\subsection{Complier average treatment effect}

Given our framework, we can see that the celebrated complier average treatment effect \citep{angrist1994,angrist1996identification} is
\begin{align*}
E[Y_1-Y_0|A_1>A_0]
=\frac{p_{01,01} - p_{01,10}}{p_{01,00} + p_{01,01} +  p_{01,10} + p_{01,11}}.
\end{align*}
Therefore, one can replace $u_j^*$ and $l_j^*$ in Algorithm~\ref{alg:upperlowerate} with 
\begin{align*}
u_j^*= \max_p \frac{p_{01,01} - p_{01,10}}{p_{01,00} + p_{01,01} +  p_{01,10} + p_{01,11}},
\end{align*}
and
\begin{align*}
l_j^*= \min_p \frac{p_{01,01} - p_{01,10}}{p_{01,00} + p_{01,01} +  p_{01,10} + p_{01,11}}.
\end{align*}
Similarly, the always-taker's, never-taker's, and defier's average treatment effect can also be obtained.

\subsection{Average treatment effect on the treated}\label{sec:G2}
Within our framework, the average treatment effect on the treated \citep{hernan2023causal} can be written as
\begin{align*}
&E[Y_1-Y_0|A_Z=1]\\
=&\frac{E[(Y_1-Y_0) I_{A_1=1}|Z=1]P(Z=1)+E[(Y_1-Y_0) I_{A_0=1}|Z=0]P(Z=0)}
{P(A_1=1|Z=1)P(Z=1)+P(A_0=1|Z=0)P(Z=0)}\\
=&[(p_{01,01}+p_{11,01}-p_{01,10}-p_{11,10})\rho+(p_{10,01}+p_{11,01}-p_{10,10}-p_{11,10})(1-\rho)]\\
&\times[(p_{01,00}+p_{01,01}+p_{01,10}+p_{01,11}+p_{11,00}+p_{11,01}+p_{11,10}+p_{11,11})\rho\\
&\ +(p_{10,00}+p_{10,01}+p_{10,10}+p_{10,11}+p_{11,00}+p_{11,01}+p_{11,10}+p_{11,11})(1-\rho)]^{-1}.
\end{align*}

\subsection{Nudge average treatment effect}

Recently, \cite{tchetgen2024nudge} proposes the so-called nudge average treatment effect, i.e., the IV estimand that recovers the average treatment effect for the subgroup of units for whom the treatment is manipulable by the instrument.
Given our framework, the nudge average treatment effect is
\begin{align*}
E[Y_1-Y_0|A_1\neq A_0]
=\frac{p_{01,01} + p_{10,01}- p_{01,10}-p_{10,10}}{p_{01,00} + p_{01,01} +  p_{01,10} + p_{01,11}+p_{10,00} + p_{10,01} +  p_{10,10} + p_{10,11}}
\end{align*}
Therefore, one can replace $u_j^*$ and $l_j^*$ in Algorithm~\ref{alg:upperlowerate} with 
\begin{align*}
u_j^*= \max_p \frac{p_{01,01} + p_{10,01}- p_{01,10}-p_{10,10}}{p_{01,00} + p_{01,01} +  p_{01,10} + p_{01,11}+p_{10,00} + p_{10,01} +  p_{10,10} + p_{10,11}},
\end{align*}
and
\begin{align*}
l_j^*= \min_p \frac{p_{01,01} + p_{10,01}- p_{01,10}-p_{10,10}}{p_{01,00} + p_{01,01} +  p_{01,10} + p_{01,11}+p_{10,00} + p_{10,01} +  p_{10,10} + p_{10,11}}.
\end{align*}

\section{Other causal assumptions}\label{sec:assumption}
\subsection{Monotonicity}\label{subsec:mono}
Monotonicity \citep{angrist1994,angrist1996identification} is a common assumption in IV analysis that ensures the instrument affects treatment assignment in a consistent direction across all units. This assumption is important in settings like complier average treatment effect identification, where we are concerned with the average treatment effect among compliers. 

Given our framework, monotonicity refers to 
$p_{10,00}=0,p_{10,01}=0,p_{10,10}=0,p_{10,11}=0$, i.e., there are no defiers.
We next show that under monotonicity and $P(A_1>A_0)>0$, the complier average treatment effect is indeed point-identified through our framework. 
\begin{proof}
When there is no defier, $p_{10,00}=0,p_{10,01}=0,p_{10,10}=0,p_{10,11}=0$.
Recall that we have 
\begin{equation*}
\begin{aligned}
    q_{000} &= p_{00,00} + p_{00,01} +  p_{01,00} + p_{01,01} \\
    q_{001} &= p_{00,10} + p_{00,11} + p_{01,10} + p_{01,11} \\
    q_{010} &= p_{10,00} + p_{10,10} + p_{11,00} + p_{11,10} \\
    q_{011} &= p_{10,01} + p_{10,11} + p_{11,01} + p_{11,11} \\
    q_{100} &= p_{00,00} + p_{00,01} + p_{10,00} + p_{10,01}\\
    q_{101} &= p_{00,10} + p_{00,11} + p_{10,10} + p_{10,11}\\
    q_{110} &= p_{01,00} + p_{01,10} +  p_{11,00} + p_{11,10} \\
    q_{111} &=  p_{01,01} + p_{01,11} + p_{11,01} + p_{11,11},
\end{aligned}
\end{equation*}
and the complier average treatment effect given $P(A_1>A_0)>0$ is 
\begin{align*}
&E[Y_1-Y_0|A_1>A_0]\\
=&\frac{p_{01,01} - p_{01,10}}{p_{01,00} + p_{01,01} +  p_{01,10} + p_{01,11}}.
\end{align*}
By a simple calculation, we have 
\begin{align*}
p_{01,01} - p_{01,10}
=q_{000}-q_{110}-q_{100}+q_{010},
\end{align*}
and
\begin{align*}
p_{01,00} + p_{01,01} +  p_{01,10} + p_{01,11}
=(q_{000}+q_{110}+q_{001}+q_{111}-q_{100}-q_{010}-q_{101}-q_{011})/2,
\end{align*}
which completes the proof.
\end{proof}


\subsection{New drug assumption}

Given our framework, the new drug assumption $A_0=0$ \citep{hernan2023causal} implies 
$p_{11,00}=0,p_{11,01}=0,p_{11,10}=0,p_{11,11}=0,p_{10,00}=0,p_{10,01}=0,p_{10,10}=0,p_{10,11}=0$, i.e., there are neither always-takers nor defiers.
We show that under the new drug assumption, the average effect on the treated is indeed point-identified through our framework. 
\begin{proof}
By applying the zero constraints $p_{11,yz} = p_{10,yz} = 0$ for all $y,z \in \{0,1\}$, we express the numerator and denominator using the given definitions of $q_{ijk}$.
For the denominator, we have
    \[
    q_{110} + q_{111} = (p_{01,10} + p_{01,00}) + (p_{01,11} + p_{01,01}) = p_{01,00} + p_{01,01} + p_{01,10} + p_{01,11}.
    \]
For the numerator, note that
    \[
    q_{000} - q_{100} = (p_{00,01} + p_{00,00} + p_{01,01} + p_{01,00}) - (p_{00,01} + p_{00,00}) = p_{01,01} + p_{01,00}.
    \]
    Subtracting $q_{110} = p_{01,10} + p_{01,00}$ then isolates the targeted difference:
    \[
    (q_{000} - q_{100}) - q_{110} = (p_{01,01} + p_{01,00}) - (p_{01,10} + p_{01,00}) = p_{01,01} - p_{01,10}.
    \]

Therefore, we have the average effect on the treated in the desired form in terms of $q_{ijk}$:
$$\frac{q_{000} - q_{100} - q_{110}}{q_{110} + q_{111}}.$$
\end{proof}

\end{document}